\documentclass[a4paper,fleqn,usenatbib]{mnras}

\usepackage{newtxtext,newtxmath}

\usepackage[T1]{fontenc}
\usepackage{ae,aecompl}

\usepackage{graphicx}	
\usepackage{amsmath}	
\usepackage{amssymb}	
\usepackage{lipsum}
\usepackage{float}

\newcommand{\mrm}[1]{\mathrm{#1}}	
\newcommand{\DRSun}[1]{$\mrm{DR_{\odot}}$} 
\newcommand{\ERSun}[1]{$\mrm{ER_{\odot}}$} 
\newcommand{\MFSun}[1]{$\mrm{MF_{\odot}}$} 
\newcommand{\ellsum}[1]{\ell_{\Sigma}} 

\title[Connecting the large- and the small-scale magnetic fields]{Connecting the  large- and the small-scale magnetic fields of solar-like stars}

\author[L. T. Lehmann et al.]{
L. T. Lehmann,$^{1}$\thanks{E-mail: ltl@st-andrews.ac.uk}
M. M. Jardine,$^{1}$
D. H. Mackay,$^{2}$
A. A. Vidotto$^{3}$
\\
$^{1}$SUPA, School of Physics and Astronomy, University of St Andrews, St Andrews KY16 9SS, UK \\
$^{2}$School of Mathematics and Statistics, University of St Andrews, St Andrews KY16 9SS, UK\\
$^{3}$School of Physics, Trinity College Dublin, The University of Dublin, Dublin-2, Ireland\\
}

\date{Accepted XXX. Received YYY; in original form ZZZ}

\pubyear{2017}

\begin{document}
\label{firstpage}
\pagerange{\pageref{firstpage}--\pageref{lastpage}}
\maketitle

\begin{abstract}

A key question in understanding the observed magnetic field topologies of cool stars is the link between the small- and the large-scale magnetic field and the influence of the stellar parameters on the magnetic field topology. We examine various simulated stars to connect the small-scale with the observable large-scale field. The highly resolved 3D simulations we used couple a flux transport model with a non-potential coronal model using a magnetofrictional technique. The surface magnetic field of these simulations is decomposed into spherical harmonics which enables us to analyse the magnetic field topologies on a wide range of length scales and to filter the large-scale magnetic field for a direct comparison with the observations. 
We show that the large-scale field of the self-consistent simulations fits the observed solar-like stars and is mainly set up by the global dipolar field and the large-scale properties of the flux pattern, e.g. the averaged latitudinal position of the emerging small-scale field and its global polarity pattern. The stellar parameters flux emergence rate, differential rotation and meridional flow affect the large-scale magnetic field topology. An increased flux emergence rate increases the magnetic flux in all field components and an increased differential rotation increases the toroidal field fraction by decreasing the poloidal field. The meridional flow affects the distribution of the magnetic energy across the spherical harmonic modes.

\end{abstract}

\begin{keywords}
stars: activity -- stars: magnetic field -- stars: solar-type -- methods: analytical
\end{keywords}



\section{Introduction}

Surveys, e.g. MagIcS\footnote{http://www.ast.obs-mip.fr/users/donati/magics/v1/}, Bcool\footnote{http://bcool.ast.obs-mip.fr/Bcool/Bcool\_\_\_cool\_magnetic\_stars.html}, MaTYSSE\footnote{https://matysse.irap.omp.eu/doku.php}, Toupies\footnote{http://ipag.osug.fr/Anr\_Toupies/}, have uncovered several trends of the large-scale magnetic field topology with stellar parameters. To understand the various observed stellar magnetic field topologies and their dependencies it is recommended to understand the solar magnetic field topology first. 

The first dynamo model for the Sun was introduced by \cite{Parker1955} after \cite{Larmor1919} suggested that the solar magnetic field may be induced by plasma motion. Many different dynamo models were constructed afterwards, for example the flux transport model by \cite{Babcock1961} and \cite{Leighton1969}, see also review by \cite{Mackay2012}. The flux transport model reproduces many of the observed solar behaviours by injecting bipolar spot pairs into the photosphere which are then affected by the surface flux transport processes: differential rotation, meridional flow and diffusion \citep{Wang1989, Baumann2004, Mackay2004, Jiang2013}. 
The flux transport models are often complemented by further approaches: e.g. 1) a 3D model, where the source of poloidal field is explicitely connected to the emerging bipoles \citep{Miesch2016, Karak2017} or 2) combining a thin-layer $\alpha\Omega$-dynamo with buoyancy instabilities for 3D flux tubes rises and a horizontal flux transport at the surface \citep{Isik2011} or 3) the connection of the photospheric flux transport model with a non-potential coronal evolution model using a magnetofrictional technique \citep{Gibb2016}. The importance of the flux transport processes for 3D solar dynamo models was recently highlighted by \cite{Cameron2015}.

Differential rotation describes the effect that many stars rotate faster on the equator than on the poles. The field lines are dragged and furled around the star and the surface magnetic field features are elongated in East-West direction. This induces a shear in the corona and makes the corona more non-potential \citep{VanBallegooijen2000, Mackay2006, Gibb2016}. The meridional flow is a poleward flow that drags the magnetic features to higher latitudes \citep{Babcock1961, DeVore1985} and is one of the key parameters for flux transport models to simulate the observed polar spot configurations, e.g. \cite{Schrijver2001}.
Additionally, the granulation and supergranulation breaks up magnetic features and is often described by diffusion processes \citep{Leighton1964}. The flux emergence rate determines how often bipolar spot pairs and magnetic features appear at the stellar surface. They result from magnetically buoyant flux tubes rising through the convection zone, e.g. \cite{Fan2001, Holzwarth2007, Weber2016}. The solar flux emergence pattern is characterised by bipolar sunspot pairs that appear between $\pm35^{\circ}$ latitudinal range \citep{Priest1982}. The bipoles are predominantly tilted  towards the equator and show the opposite polarities on the different hemispheres \citep{Hale1919}. 

\cite{Gibb2016} explored the influence of the flux emergence rate and differential rotation on the non-potential coronal field using a flux transport model based on solar observations in connection with a non-potential coronal evolution model \citep{Mackay2006, Yeates2012}. They found that an increased flux emergence rate adds more flux to the corona but leaves the global corona structure unchanged whilst an increased differential rotation opens the corona up and makes it more non-potential.

The first detection of a magnetic field on a solar-like star was undertaken by \cite{Robinson1980}. The total magnetic flux at the stellar surface is measurable by the Zeeman broadening in spectral lines but there is little to no ability to uncover the magnetic field topology \citep{Robinson1980, Saar1988, Reiners2006, Lehmann2015, Scalia2017}. The Zeeman Doppler Imaging (ZDI) technique \citep{Semel1989, Donati1997, Donati2006} analyses time series of polarised light profiles and provides the large-scale magnetic field topology in intensity and orientation. ZDI suffers from cancellation effects due to the cancellation of  opposite polarities on smaller scale \citep{Johnstone2010, Arzoumanian2011, Lang2014}. As a result ZDI is only able to observe the large-scale magnetic field topology and the resolution is dependent, among other parameters, on the $v \sin i$ of the star, e.g. \cite{Morin2010}. 

Several surveys have uncovered the magnetic field topologies of stars with different masses, ages and environments, e.g. \cite{Donati2006a, Marsden2006, Petit2008, Morin2010, Fares2013, Folsom2016, Hebrard2016, Hill2017}. Stellar activity phenomena driven by the magnetic field scale up with rotation and earlier spectral types, e.g. \cite{Skumanich1972, Hartmann1987, Guedel2007, Reiners2012}. With shorter rotation periods the x-ray luminosity increases  \citep{Pallavicini1981, Walter1981}, the choromopheric emission \citep{Middelkoop1981, Mekkaden1985} and the mean magnetic field \citep{Vidotto2014, Folsom2016}. The stellar activity is also dependent on the stellar mass. Stars owning the same rotation period display an increased activity with decreasing mass \citep{Donati2009, Marsden2014}. 

\cite{Petit2008} showed that not only does the magnetic field strength increase with rotation but also the magnetic field topology changes as the fraction of toroidal field increases with rotation. Additionally, \cite{See2015} discovered that the toroidal field scales more steeply with inverse Rossby number (which is the ratio between the rotation period and the turnover time in the convection cell) then the poloidal field. They also found two powerlaw dependencies for different mass ranges between the toroidal and poloidal magnetic field energies. Cool stars with masses above half a solar mass show a steeper dependence of $\langle B^2_{\mathrm{tor}}\rangle \propto \langle B^2_{\mathrm{pol}}\rangle^{1.25\pm0.06}$ compared to the low mass stars with $\langle B^2_{\mathrm{tor}}\rangle \propto \langle B^2_{\mathrm{pol}}\rangle^{0.72\pm0.08}$. It was also revealed that stars with higher fractions of toroidal field show  higher fractions of axisymmetric fields.

The large-scale magnetic topologies of the cool stars can be generally summarised by three groups: the stars with Rossby numbers greater than one, e.g. the Sun and 61~Cyg~A, show weak, mainly poloidal and axisymmetric fields \citep{Donati2009}. The faster rotating stars with Rossby numbers less than one and masses above $0.5\,\mrm{M_{\odot}}$, e.g. EK Dra and DS Leo, show strong to dominant toroidal fields, non-axisymmetric poloidal fields and a higher field complexity, \cite{Donati2008}. The low mass stars with masses below $0.5\,\mrm{M_{\odot}}$, e.g. WX Uma and DX Cnc, show either very strong, simple, axisymmetric poloidal fields or weaker, complexer and more toroidal fields \citep{Morin2008a, Morin2010}.

The different magnetic field topologies might suggest that there are also different flux emergence pattern and rates. Estimating the flux emergence pattern starts to become possible by modelling the star spot occulations of transiting planets, e.g. \cite{Morris2017}. The meridional flow is unknown for cool stars other than the Sun but could be estimated by the Hale cycle \citep{Baklanova2015}. Furthermore, \cite{Hung2017} used a data assimilation method based on solar observations and an axisymmetric mean field dynamo model to estimate the meridional circulation for the Sun. In contrast differential rotation is reliably detectable by the modulation of chromopheric emission lines \citep{Donahue1996} or more precisely with ZDI \citep{Donati1997, Petit2002, Marsden2006, Waite2011, Marsden2011}. It was found that differential rotation increases with higher effective temperature and stellar mass \citep{Barnes2005, Cameron2007, Kueker2011}. 

In this paper, we want to investigate the influence of the stellar parameters: flux emergence rate, differential rotation and meridional flow on the large- and small-scale magnetic field topology using 3D non-potential simulations based on \cite{Gibb2016}. The aim is to connect the small- with the large-scale field to understand the observed magnetic field topologies and their dependencies. We are not re-analysing the results of \cite{Gibb2016}, which focus on the coronal response. We concentrate on the photosphere of their self-consistent fully 3D simulations as the magnetic field topology is usually observed at this layer. Furthermore, we want to improve the accessibility of the simulations to the observers and provide the decomposition of the surface magnetic field topology into the magnetic field components radial, azimuthal and meridional and poloidal and toroidal, which are widely used by the observers.

The paper is structured as follows: Section~\ref{sec:ModellingTechniques} describes the applied modelling techniques and Section~\ref{sec:Simulations} the flux transport simulations used. Our results are presented in Section~\ref{sec:Results} and discussed in Section~\ref{sec:Discussion}. Our conclusions and a summary of the main results is provided in Section~\ref{sec:Summary}.

\section{Modelling Techniques}
\label{sec:ModellingTechniques}

The main drawback for a direct comparison between simulated and observed stellar magnetic field vector maps is their immense difference in resolution. The resolution of the simulated stellar magnetic field maps allows very high resolution in the sub-degree regime. The resolution of the observed stellar magnetic field maps is relatively low for slow-rotating solar-like stars and depends on many factors such as stellar $v\sin i$, phase coverage, data quality and more. A direct comparison between the simulations and the observations is not possible unless one adjusts the resolution of the higher resolved simulations to the resolution of the lower resolved observations.

\subsection{The different magnetic field components and their decomposition into their spherical harmonics}

By decomposing the simulated magnetic field vector maps into their spherical harmonics one can filter the large- and the small-scale field by selecting the corresponding spherical harmonic modes~$\ell$. The selected length scale is approximately described by $\theta \approx 180^{\circ}/\ell$. By selecting the lower spherical harmonic modes of the simulated magnetic field map, e.g. $\ell \le 5$ or $\ell \le 10$, we select the corresponding large-scale field of the simulated magnetic field topology, see also Figure~1, \cite{Lehmann2017}. This allows a fair order of magnitude comparison between the large-scale field of the simulated magnetic field vector maps with most of the observed stellar magnetic field vector maps. Filtering the large-scale field by selecting the low spherical harmonic modes was previously shown by e.g. \cite{Morin2010}, \cite{Johnstone2014}, \cite{Yadav2015}, \cite{Vidotto2016}, \cite{Folsom2016}, \cite{Lehmann2017}.

The stellar magnetic field topology is often described by the radial, azimuthal and meridional component, e.g. \cite{Morin2008a, Morin2010, Fares2012, Rosen2016}, or by the poloidal and toroidal component, e.g. \cite{Petit2008, Donati2009, See2015, Vidotto2016}. Additionally, the axisymmetric component is used to express how much an individual magnetic field component or mode is aligned with the rotation axis. 

The poloidal and toroidal component can be defined as a composition of several spherical harmonic modes of the three magnetic field vectors $B_r, B_{\theta}, B_{\phi}$ after \cite{Elsasser1946} and \citet[Appendix III]{Chandrasekhar1961}. \cite{Donati2006a}, Eq.~(2)-(8), characterises the poloidal field by the coefficients $\alpha_{\ell m}$ and $\beta_{\ell m}$ and the toroidal field by $\gamma_{\ell m}$.

 \begin{align}
 B_{\mathrm{pol}, r}(\theta, \phi) &\equiv B_{r}(\theta, \phi) = \sum_{\ell m} \alpha_{\ell m} P_{\ell m} e^{im\phi}, \\
 B_{\mathrm{pol},\theta}(\theta,\phi) &= \sum_{\ell m} \beta_{\ell m} \frac{1}{\ell+1} \frac{\mathrm{d}P_{\ell m}}{\mathrm{d}\theta} e^{im\phi}, \\
 B_{\mathrm{pol}, \phi}(\theta, \phi) &= - \sum_{\ell m} \beta_{\ell m} \frac{im P_{\ell m} e^{im\phi}}{(\ell + 1) \sin \theta},
 \end{align}

 \begin{align}
 B_{\mathrm{tor}, r}(\theta, \phi) &= 0, \\
 B_{\mathrm{tor},\theta}(\theta,\phi) &= \sum_{\ell m} \gamma_{\ell m} \frac{im P_{\ell m} e^{im\phi}}{(\ell + 1) \sin \theta}, \\
 B_{\mathrm{tor}, \phi}(\theta, \phi) &=  \sum_{\ell m} \gamma_{\ell m} \frac{1}{\ell+1} \frac{\mathrm{d}P_{\ell m}}{\mathrm{d}\theta} e^{im\phi},
 \end{align}
 so that $\vec{B}_{\mathrm{pol}} + \vec{B}_{\mathrm{tor}} = \vec{B}$. The radial field points outwards, the meridional ($\theta$) field increases with colatitude from north to south and the azimuthal field ($\phi$) increases with longitude in the direction of the rotation. $P_{\ell m} \equiv c_{\ell m}P_{\ell m}(\cos \theta)$ is the associated Legendre polynomial of mode $\ell$ and order $m$, where $c_{\ell m}$ is a normalization constant:
 \begin{equation}
 c_{\ell m} = \sqrt{\frac{2\ell+1}{4\pi}\frac{(\ell - m)!}{(\ell + m)!}}.
 \end{equation} 
 The sums run from $1\leq \ell \leq \ell_{\mathrm{max}}$ and $\vert m \vert \leq \ell$, where $\ell_{\mathrm{max}}$ is the maximum mode of the spherical harmonic decomposition. The axisymmetric modes are selected by $m=0$. Otherwise, we sum over all $m$ for the magnetic field of a given mode $\ell$. 

We decompose the simulated surface magnetic field topology into their spherical harmonic modes for the radial, azimuthal and meridional component and for the poloidal and toroidal component using the decomposition described by \cite{Vidotto2016a}. We determine the mean squared flux density for the different field components, e.g. for the poloidal field: 
\begin{equation}
\langle B^2_{\mathrm{pol}}\rangle = \tfrac{1}{4\pi}\textstyle \int \textstyle B^2_{\mathrm{pol}}(\theta, \phi)\sin(\theta)\,\mathrm{d}\theta \mathrm{d}\phi
\end{equation}
for the individual $\ell$-modes $\ell = 1-28$ and for the cumulative $\ell$-modes $\ellsum\ \leq 1-28$.
From now on we call the mean squared flux density $\langle B^2\rangle\mathrm{[G^2]}$ magnetic energy. Although the mean squared flux density is not exactly equivalent to the magnetic energy it is a good proxy for the simulations. For the observations $\langle B^2\rangle\mathrm{[G^2]}$ is restricted to the net magnetic flux of the resolution elements but widely referred as magnetic energy instead of mean squared flux density, see e.g. the review from \cite{Reiners2012}. 
The magnetic energy of the cumulative $\ellsum\ $-modes includes also the energy of all lower $\ell$-modes while the magnetic energy for an individual $\ell$-mode only includes the energy for this specific $\ell$-mode. For example the magnetic energy of the toroidal quadrupolar mode describes the energy of the toroidal field stored in the $\ell = 2$ while the magnetic energy of the toroidal cumulative quadrupolar mode describes energy of the toroidal field stored in $\ellsum\ \le 2$ including the magnetic energy of the dipolar $\ell = 1$ and the quadrupolar $\ell = 2$ mode.

\subsection{Estimating the rotation period of the simulations and the optimal average}
\label{ProtandAveraging}

The simulations based on the work of \cite{Gibb2016} model the surface magnetic field topolgy of stars with different flux emergence rates, differential rotation rates and meridional flows, which corresponds to stars of different rotation periods. To estimate the stellar rotation we use the result that the mean magnetic flux density is correlated to the stellar rotation. 
\cite{Saar1996} found that
\begin{equation}
\langle \vert B_I \vert \rangle \propto P_{\mrm{rot}}^{-1.7},
\label{Eq:Prot}
\end{equation}
where $\langle \vert B_I \vert \rangle = fB_I$ is the total mean unsigned flux density from Zeeman-Broadening measurements. \cite{Vidotto2014} showed that also the averaged large-scale field strength $\langle \vert B_I \vert \rangle$ from ZDI measurements correlates with the rotation period ($\langle \vert B_V \vert \rangle \propto P_{\mrm{rot}}^{-1.32\pm 0.14}$). The simulated surface vector magnetic field maps that we used are highly resolved, which allows us to determine the total mean flux density and to use the correlation found by \cite{Saar1996}.
We determine the total mean magnetic flux density for each day of the simulations and average them for one stellar parameter set to get $\langle \vert B_{\mrm{tot}} \vert \rangle = \langle \vert B_I \vert \rangle$. We use Eq.~\ref{Eq:Prot} to estimate the rotation period, where the solar case is defined to rotate with $P_{\mrm{rot}} = 27\,\mathrm{d}$ for $\langle \vert B_{\mrm{tot}} \vert \rangle = 14.7\,\mathrm{G}$. Table \ref{Tab:Prot} shows the rotation periods for the simulations, which range from 13.9 days to 33.4 days. Increasing the flux emergence rate results in a strong increase of the total mean magnetic flux density and therefore in a strong decrease of the rotation period.  Increasing the differential rotation or the meridional flow leads to cancellations of small-scale magnetic field structures of opposite polarity and decreases the total mean magnetic flux density which is reflected in a longer rotation period. The differential rotation has the greater effect on the inferred rotation period compared with the meridional flow.

 \begin{table}
 \caption{The rotation period in days for the simulations. The simulations vary in flux emergence rate (ER), differential rotation (DR) and meridional flow (MF), which are displayed in terms of the solar values. }
 \textbf{a.} For solar meridional flow $\mrm{MF} = \mrm{MF_{\odot}}$:\\
 \begin{tabular}{c|cccccc}
 \hline
 ER$\setminus$DR & $0.3\,\mrm{DR_{\odot}}$ & $0.5\,\mrm{DR_{\odot}}$ & $0.8\,\mrm{DR_{\odot}}$ & $1\,\mrm{DR_{\odot}}$ & $3\,\mrm{DR_{\odot}}$ & $5\,\mrm{DR_{\odot}}$ \\
 \hline
 $1\,\mrm{ER_{\odot}}$ & 25.2 & 25.8 & 26.2 & 27.0 & 30.4 & 32.0 \\
 $3\,\mrm{ER_{\odot}}$ & 16.5 & 16.6 & 17.0 & 17.4 & 19.0 & 20.2 \\
 $5\,\mrm{ER_{\odot}}$ & 13.9 & 14.1 & 14.2 & 14.3 & 15.9 & 17.0 \\
 \hline
 \end{tabular}\\
 \newline
 \textbf{b.} For higher meridional flow $\mrm{MF} = 10\,\mrm{MF_{\odot}}$:\\
  \begin{tabular}{c|ccc}
 \hline
  ER$\setminus$DR & $1\,\mrm{DR_{\odot}}$ & $3\,\mrm{DR_{\odot}}$ & $5\,\mrm{DR_{\odot}}$ \\
 \hline
  $1\,\mrm{ER_{\odot}}$  & 26.8 & 30.7 & 33.4 \\
  $3\,\mrm{ER_{\odot}}$ & 16.7 & 19.5 & 20.6 \\
  $5\,\mrm{ER_{\odot}}$ & 14.0 & 15.7 & 16.4 \\
 \hline
 \end{tabular}
 \label{Tab:Prot}
 \end{table}

The results and errors of the following sections are calculated via an optimal average algorithm, which mimics the observational restrictions. The simulations provide for every star $n_{\mrm{tot}} \gtrapprox 300$ magnetic maps for every day $j(i,m)$, 
\begin{equation}
j(i,m) = i + m \cdot  P_{\mrm{rot}}
\end{equation}
where $i \in (0, n_i)$ is the start date and $m \in (0, n_m)$ is a counter for the multiple of the rotation period $P_{\mrm{rot}}$, while $n_i = P_{\mrm{rot}}-1$ and $n_m = \mrm{mod}\left(\frac{n_{\mrm{obs}}}{P_{\mrm{rot}}}\right)$.  
For an observed stellar magnetic field map one needs a full stellar rotation to observe the whole visible surface. We calculate the arithmetic mean $\mu(i)$ for any magnetic field parameter $F\left(j(i,m)\right)$ by averaging over the multiple $m$ of the rotation period $P_{\mrm{rot}}$ from a certain start date $i$, 
\begin{equation}
\mu(i) = \frac{\sum_{m=1}^{n_m} F\left(j(i,m)\right)}{n_m}.
\end{equation}
This is equivalent to an arithmetic average over $n_m$ subsequently observed stellar magnetic field maps. We have the advantage to obtain surface magnetic field maps for every day of the simulation and not only one snapshot per stellar rotation. We calculate therefore the  optimal average $\sigma_{\mrm{opt\ avg}}$ over the arithmetic means $\mu(i)$ for the different start dates $i$, 
\begin{equation}
\sigma_{\mrm{opt\ avg}} = \frac{\sum_{i=1}^{n_i} \mu(i)}{n_i}.
\end{equation}
Further, we determine the corresponding 1-$\sigma$ standard derivation using the same concept.

\section{Simulations}
\label{sec:Simulations}

To investigate the poloidal and toroidal nature of simulated non-potential stellar magnetic fields, we analyse the 3D non-potential simulations of \cite{Gibb2016}. The simulations follow the coupled evolution of the 3D vector photospheric and coronal magnetic fields of a star out to 2.5\,$\mrm{R_{\star}}$, as the field evolves in response to surface magnetic flux transport processes. These surface process include the large-scale advection of magnetic fields due to differential rotation, meridional flow and surface diffusion. Magnetic flux emergence is also included to produce varying levels of magnetic activity. As the surface effects act on the coronal field of the star, the field evolves through a series of non-linear force-free states with the build-up of electric currents and free magnetic energy.
The simulations extend over a time period of 1 year and have a spatial resolution of 0.9375 degrees at the equator. While the full 3D field is available out to 2.5$\,\mrm{R_{\star}}$, in the present paper we focus on the non-potential vector magnetic field ($B_r, B_{\theta}, B_{\phi}$) that is produced at the photosphere and decompose it into its poloidal and toroidal components to provide a better comparison with observations.
 
Since stellar flux emergence profiles and rates are currently unknown, the properties of emerging bipoles are based on solar emergence profiles where the frequency, location and times of emergence are described through a parameterised emergence model described in Section 3 of \cite{Gibb2016}.  A variety of simulations are carried out where we investigate a number of possible combinations of parameters. These include varying the flux emergence rate, which is set to be 1, 3 and 5 times the solar flux emergence rate (deduced during the maximum of solar cycle 23). Correspondingly, the differential rotation rate varies between 0.3-5 times that observed for the Sun. Finally, the meridional flow rate is considered for both solar values and 10 times the solar rate, which is motivated by the results of \cite{Mackay2004}. They showed that an increase of the meridional flow by a factor of 10 is required to explain the observed high-latitude intermingled flux of faster rotating stars.
As described in \cite{Gibb2016} all other parameters and profiles within the surface flux transport model are set to solar values as currently their profiles and rates are unknown for stars. Full details of the simulations can be found in Section 2 and 3 of \cite{Gibb2016}.

\section{The magnetic field topologies of the simulated stars}
\label{sec:Results}

To investigate the magnetic field properties of the simulated vector magnetic field maps, we analyse their surface magnetic field topology. In particular, we consider the energy of the poloidal and toroidal magnetic field components (Section~\ref{subsec:SimVsObs} to \ref{subsec:MSFD}), the energy distribution across the spherical harmonic $\ell$-modes (Section~\ref{subsec:Distribution}), and the axisymmetry of the field (Section~\ref{subsec:Axisymmetry}). We compare our results with the analysis of observed stars published by \cite{See2015}, where possible.
Furthermore, we identify dependencies of the simulated surface magnetic field topology on the stellar properties: flux emergence rate, differential rotation and meridional flow.

\subsection{Comparing the simulations with observations}
\label{subsec:SimVsObs}

\begin{figure}[h]
	\includegraphics[width=\columnwidth]{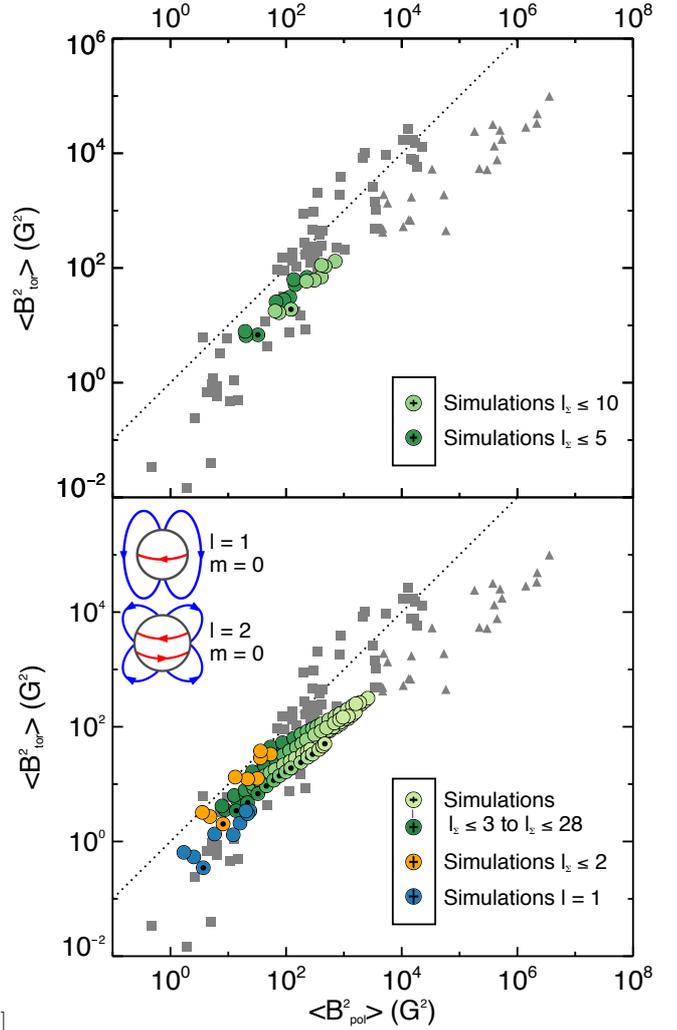}
    \caption{The magnetic energy stored in the poloidal $\langle B^2_{\mathrm{pol}}\rangle$ and the toroidal component $\langle B^2_{\mathrm{tor}}\rangle$. The observations are displayed by grey symbols, where star with stellar masses $M_{\star} \ge 0.5\,\mrm{M_{\odot}}$ are plotted as squares and stars with masses $M_{\star} < 0.5\,\mrm{M_{\odot}}$ as triangles. The simulations covering flux emergence rates of $\mrm{ER} = 1,3,5\,\mrm{ER_{\odot}}$ and a differential rotation of $\mrm{DR} = 1,3,5\,\mrm{DR_{\odot}}$ are shown as coloured circles. The simulated Sun is indicated by the solar symbol $\odot$ and the dashed line indicates equal poloidal and toroidal energies. \textit{Top:} For a direct comparison between the simulations and observations the simulations are restricted to the large-scale field by spherical harmonics up to $\ellsum\  \leq 5$ (dark green circles) or $\ellsum\  \leq 10$ (light green circles). \textit{Bottom:} Including all surface scale sizes for the simulations: the dipolar mode $\ell=1$ (blue circles), the cumulative quadrupolar mode $\ellsum\  \leq 2$ (orange circles), and the higher cumulative $\ellsum\ $-modes $\ellsum\  \leq 3$ to $\ellsum\  \leq 28$ (greenish circles), where the colour gets lighter with increasing $\ellsum\ $-modes. The higher $\ellsum\ $-modes follow the powerlaw $\langle B^2_{\mathrm{tor}}\rangle \propto \langle B^2_{\mathrm{pol}}\rangle^{0.77\pm0.02}$. The inserts show the poloidal (blue) and toroidal (red) field lines for the axisymmetric dipole and quadrupole mode. This Figure uses a similar format to that in \citet{Lehmann2017}, Fig.~2.
		}
    \label{fig:BtorBpolRepeatori}
\end{figure}

We first investigate the magnetic energy budgets of the poloidal and toroidal fields, see Fig.~\ref{fig:BtorBpolRepeatori}.
We use a similar format to that in \cite{Lehmann2017}, Fig.~2. Figure 2 of \cite{Lehmann2017}, showed only $\langle B^2\rangle$ of a single vector magnetic field map per simulated star. In contrast Figure~\ref{fig:BtorBpolRepeatori} in this paper displays the optimal average over more than 300 vector magnetic field maps per simulated star. We present the results for nine different stars with flux emergence rates of $\mrm{ER} = 1,3,5\,\mrm{ER_{\odot}}$ and a differential rotation of $\mrm{DR} = 1,3,5\,\mrm{DR_{\odot}}$ in Fig.~\ref{fig:BtorBpolRepeatori}. They cover the same parameter range as the simulations shown in Fig.~2, \cite{Lehmann2017}, and allow a direct comparison. Figure~\ref{fig:BtorBpol_All} in the appendix includes all of the 27 simulated stars available in this study using the same format as Fig.~\ref{fig:BtorBpolRepeatori}.

The grey symbols in Figure~\ref{fig:BtorBpolRepeatori} represent the results published in Fig.~2 \textit{top} of \cite{See2015}, for a sample of 55 observed cool stars\footnote{The observations including results from the Bcool and Toupies survey were published by Petit (in preparation); \cite{BoroSaikia2015, doNascimento2014, Donati2003, Donati2008, Fares2009, Fares2010, Fares2012, Fares2013, Folsom2016, Morin2008a, Morin2008, Morin2010, Jeffers2014,  Petit2008, Waite2011}.}, where the stars with masses above $0.5\,\mrm{M_{\odot}}$ are displayed by squares and less by triangles. The \textit{top} panel of Fig.~\ref{fig:BtorBpolRepeatori} shows the magnetic energy $\langle B^2\rangle$ for the simulations when including the cumulative spherical harmonic $\ellsum\ $-modes $\ellsum\  \le 5$ (darker green circles) and $\ellsum\  \le 10$ (lighter green circles). These $\ellsum\ $-modes mimic two typical resolutions of the observed stars for slowly and moderately rotating stars and allow the direct comparison of the simulations with the observations. Stellar magnetic field reconstructions based on observations often include only $\ellsum\  \leq 5$ (e.g. \citealt{Morin2010}, \citealt{Vidotto2016a}, \cite{Folsom2016}) or $\ellsum\  \leq 10$ modes (e.g. \citealt{Johnstone2014}, \citealt{Yadav2015}), where high-resolution solar synoptic maps reach, e.g., $\ellsum\  \le 192$ \citep{DeRosa2012}. Restricting the highly resolved simulations to low $\ellsum\ $-modes allows an unbiased comparison between simulations and observations. The simulation representing the solar case is marked with the solar symbol $\odot$. The dashed line indicates the unity line, where the poloidal and toroidal energies are equal.  
We see in Fig.~\ref{fig:BtorBpolRepeatori} \textit{top}, that the simulations (coloured circles) are within the regime of the observations (grey symbols). To be more precise the simulations cover the same parameter space as solar-like stars, e.g. HN\,Peg \citep{BoroSaikia2015} or $\varepsilon$\,Eri \citep{Jeffers2014}.

The \textit{bottom} panel of Figure~\ref{fig:BtorBpolRepeatori} includes all of the cumulative $\ellsum\ $-modes up to $\ellsum\  \le 28$ overplotting the observations. One cannot compare the different $\ellsum\ $-modes directly with the observations as they have different resolutions. For a direct comparison see Fig.~\ref{fig:BtorBpolRepeatori} \textit{top}. We colour-code the cumulative $\ellsum\ $-modes in three regimes: the dipolar modes (blue), the quadrupolar modes (orange) and the higher $\ellsum\ $-modes (varying green colour), where the colour becomes lighter with increasing $\ellsum\ $-modes. The dipolar modes are mainly poloidal. The quadrupolar modes show often one order of magnitude higher toroidal fields and are the modes with the highest fraction of toroidal field. The inserts in the \textit{bottom} panel of Fig.~\ref{fig:BtorBpolRepeatori} show the poloidal (blue) and toroidal (red) field of an axisymmetric dipole and quadrupole. The higher $\ellsum\ $-modes show a fixed ratio between the toroidal and poloidal field of $\langle B^2_{\mathrm{tor}}\rangle \propto \langle B^2_{\mathrm{pol}}\rangle^{0.77\pm0.02}$. This powerlaw is remarkably similar to the powerlaw for the low mass M-dwarfs (grey triangles) $\langle B^2_{\mathrm{tor}}\rangle \propto \langle B^2_{\mathrm{pol}}\rangle^{0.72\pm0.08}$ found by \cite{See2015}, see also discussion in Section~\ref{sec:Discussion}.

Comparing the optimal average of the $\approx$ 300 simulated vector magnetic maps per star in Fig.~\ref{fig:BtorBpolRepeatori} with the single simulated vector magnetic map per star in Fig.~2 \citep{Lehmann2017}, we notice that the optimal average reduces the spread and the trends for the $\ellsum\ $-modes become clearer, especially for the higher $\ellsum\ $-modes. Both cover a similar parameter space in $\langle B^2_{\mathrm{pol}}\rangle$ and $\langle B^2_{\mathrm{tor}}\rangle$ and are within the range set by observations of solar-like stars. The solar-case simulation in Fig.~\ref{fig:BtorBpolRepeatori} is now closer to the other simulations compared to Fig.~2 in \cite{Lehmann2017}. This confirms the suggestion made in \cite{Lehmann2017} that the analysed simulated solar magnetic field map had a generally lower toroidal field than the average for the solar-case simulations.

Furthermore, we calculate the 1-$\sigma$ standard deviations of the optimal average. The errors are most of the time equal or smaller than the plot symbol. We display the maximal errorbars for the shown $\ellsum\ $-modes in the legends of Fig.~\ref{fig:BtorBpolRepeatori}. The error becomes smaller with increasing $\ellsum\ $-modes on the log-log scale.

To summarise, the large-scale field of the simulations match the observations of solar-like stars regarding the poloidal and toroidal energy. We notice three different behaviours for the cumulative $\ellsum\ $-modes: a mainly poloidal dipolar component, a quadrupolar component storing the highest fraction of toroidal field and the
higher $\ellsum\ $-modes following a powerlaw $\langle B^2_{\mathrm{tor}}\rangle \propto \langle B^2_{\mathrm{pol}}\rangle^{0.77\pm0.02}$.

\subsection{The impact of differential rotation and meridional flow}

\begin{figure*} 
	\includegraphics[width=2\columnwidth]{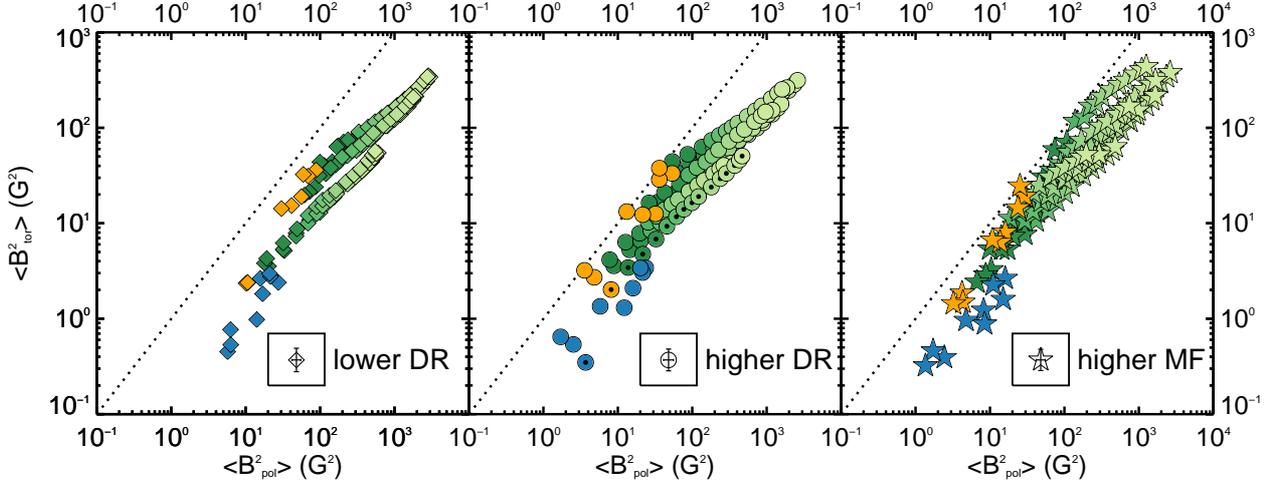}
    \caption{The poloidal $\langle B^2_{\mathrm{pol}}\rangle$ and toroidal energy $\langle B^2_{\mathrm{tor}}\rangle$ for the simulations split in three groups. All three groups cover flux emergence rates of $\mrm{ER} = 1, 3, 5$\,\ERSun\ . The same format as in Fig.~\ref{fig:BtorBpolRepeatori} \textit{bottom} is used. \textit{Left:} The lower DR simulations (diamonds) with differential rotation of $\mrm{DR} = 0.3, 0.5, 0.8$\,\DRSun\ , where the higher $\ellsum\ $-modes follow the powerlaw $\langle B^2_{\mathrm{tor}}\rangle \propto \langle B^2_{\mathrm{pol}}\rangle^{0.87\pm0.02}$. \textit{Middle:} The higher DR simulations (circles) with $\mrm{DR} = 1, 3, 5$\,\DRSun\ , where the higher $\ellsum\ $-modes follow the powerlaw $\langle B^2_{\mathrm{tor}}\rangle \propto \langle B^2_{\mathrm{pol}}\rangle^{0.77\pm0.02}$. These are the same simulation as plotted in Fig.~\ref{fig:BtorBpolRepeatori} \textit{bottom}. \textit{Right:} The higher MF simulations (stars) owning the same differential rotation as the higher DR simulation but a higher meridional flow of  $\mrm{MF} = 10\,\mrm{MF_{\odot}}$. The higher $\ellsum\ $-modes follow the powerlaw $\langle B^2_{\mathrm{tor}}\rangle \propto \langle B^2_{\mathrm{pol}}\rangle^{0.80\pm0.02}$.}
    \label{fig:BtorBpolZoom}
\end{figure*}

We now present the results of an extended parameter study for variations in the stellar properties of differential rotation and meridional flow. We split the simulations into three groups: 
\begin{itemize}
\item \textbf{lower DR} includes the simulations with sub-solar differential rotation of $\mrm{DR} = 0.3, 0.5, 0.8\,\mrm{DR_{\odot}}$, 
\item \textbf{higher DR} includes the simulations with differential rotation of $\mrm{DR} = 1, 3, 5\,\mrm{DR_{\odot}}$ and
\item \textbf{higher MF} includes the simulations with the same differential rotation as the higher DR group, but having a meridional flow ten times higher than that found on the Sun, $\mrm{MF} = 10\,\mrm{MF_{\odot}}$.
\end{itemize}
All three groups cover the flux emergence rates $\mrm{ER} = 1, 3, 5\,\mrm{ER_{\odot}}$. Figure~\ref{fig:BtorBpolZoom} compares the three groups regarding the poloidal and toroidal energy using the same format as Fig.~\ref{fig:BtorBpolRepeatori} \textit{bottom}. 

Comparing the group of lower DR (\textit{left} panel, diamonds) with higher DR (\textit{middle} panel, circles) we notice, that the higher DR simulations are generally closer to the dashed unity line. The poloidal energy $\langle B^2_{\mathrm{pol}}\rangle$ decreases with higher differential rotation. For the lower DR all $\ellsum\ $-modes are poloidal dominated but for the higher DR some $\ellsum\ $-modes especially the quadrupolar modes (orange circles) are close or slightly above the unity line indicating a more dominant toroidal field as $\langle B^2_{\mathrm{pol}}\rangle$ decreases while $\langle B^2_{\mathrm{tor}}\rangle$ remains widely constant. This is reflected in the shallower slope for the higher $\ellsum\ $-modes (greenish symbols) of the higher DR simulations $\langle B^2_{\mathrm{tor}}\rangle \propto \langle B^2_{\mathrm{pol}}\rangle^{0.77\pm0.02}$ compared to the lower DR simulations $\langle B^2_{\mathrm{tor}}\rangle \propto \langle B^2_{\mathrm{pol}}\rangle^{0.87\pm0.02}$.

Comparing the group of solar with higher meridional flow (\textit{middle} and \textit{right} panel) we see a clear effect for the $\ellsum\ $-modes. The simulations with solar MF and higher DR (\textit{middle} panel, circles) display three regimes for the $\ellsum\ $-modes: the dipolar modes are mainly poloidal, the quadrupolar modes show the highest fraction of toroidal field and the higher $\ellsum\ $-modes follow a fixed ratio between the poloidal and toroidal field component. An increase in the meridional flow (\textit{right} panel, star symbols) changes this clear separation into three regimes. Next to the quadrupolar mode also other low $\ellsum\ $-modes are close to the unity line. This indicates that several low to mid $\ellsum\ $-modes become stronger toroidal modes by losing $\langle B^2_{\mathrm{pol}}\rangle$ and gaining a small amount of $\langle B^2_{\mathrm{tor}}\rangle$. This causes the more curved appearance of the $\ellsum\  \ge 3$-modes. The dipolar modes appear at slightly lower $\langle B^2_{\mathrm{tor}}\rangle$ and the higher $\ellsum\ $-modes follow a powerlaw again $\langle B^2_{\mathrm{tor}}\rangle \propto \langle B^2_{\mathrm{pol}}\rangle^{0.80\pm0.02}$ .

To summarise the differential rotation seems to decrease the magnetic energy of the poloidal field $\langle B^2_{\mathrm{pol}}\rangle$ while the energy of the toroidal field $\langle B^2_{\mathrm{tor}}\rangle$ widely remains the same. The meridional flow causes a change in the large-scale field topology by affecting $\langle B^2_{\mathrm{pol}}\rangle$ and $\langle B^2_{\mathrm{tor}}\rangle$ differently for the low $\ellsum\ $-modes.

\subsection{The impact of emergence rate}
\label{subsec:MSFD}

\begin{figure} 
	\includegraphics[width=\columnwidth]{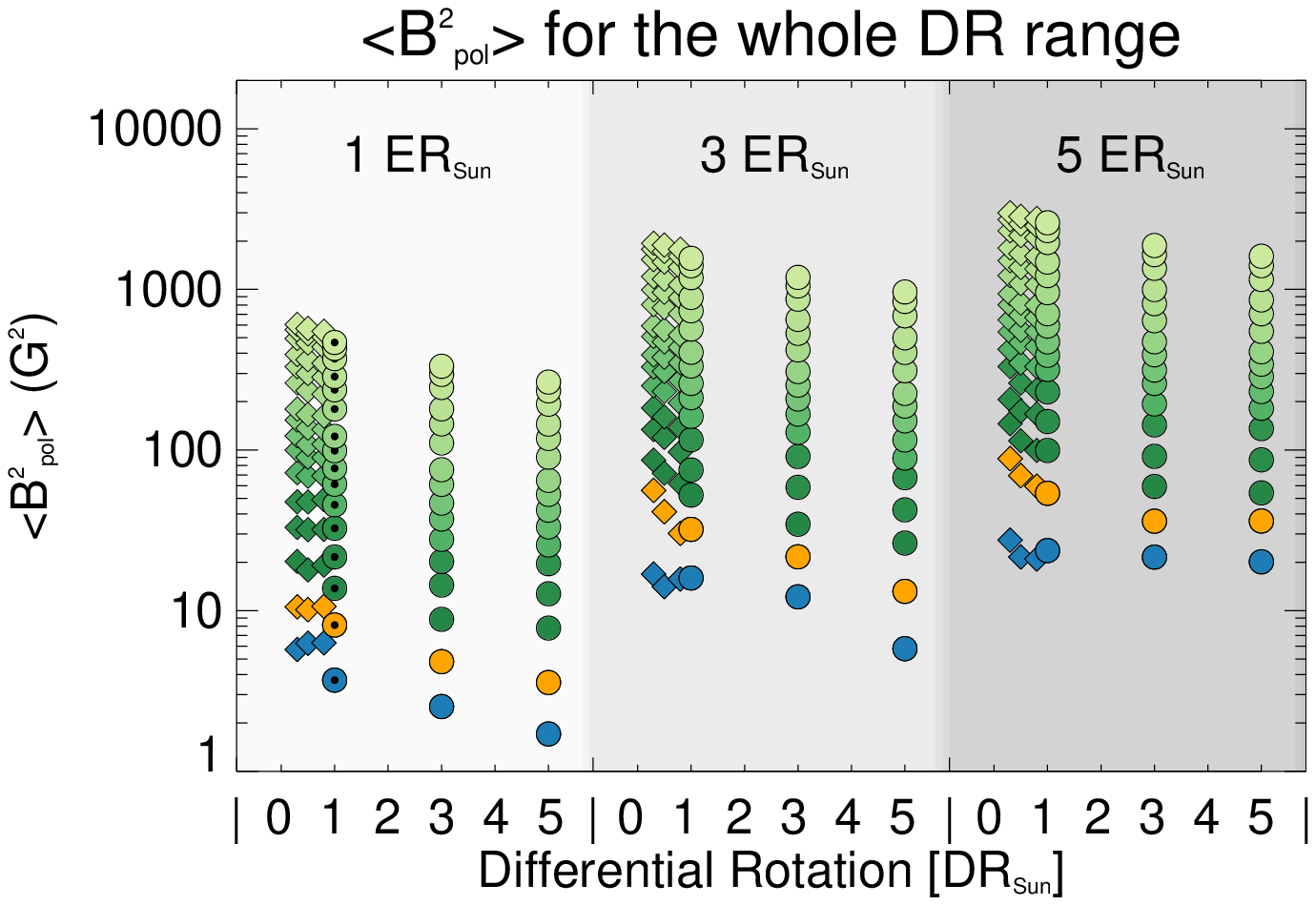}
	\includegraphics[width=\columnwidth]{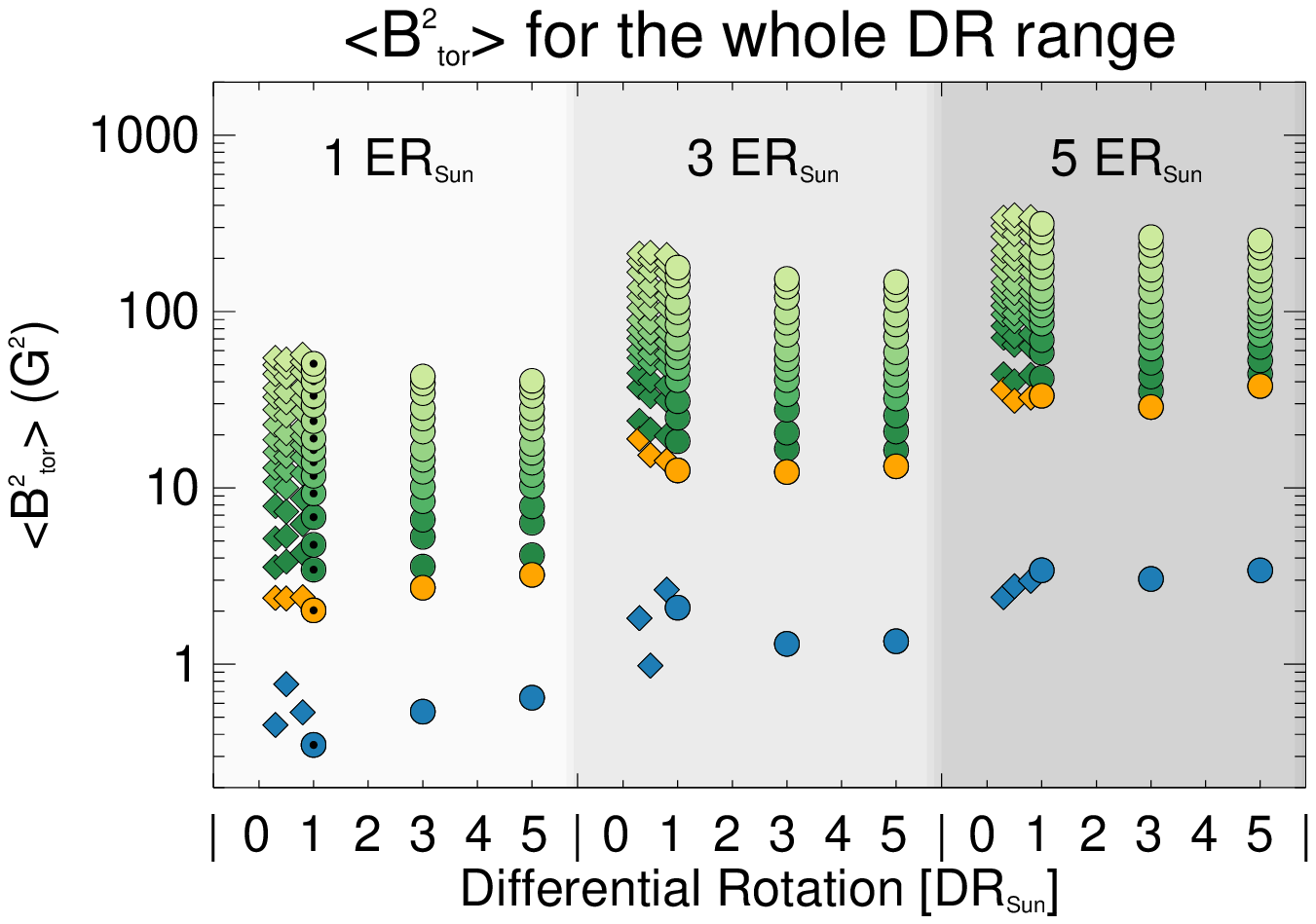}
   \caption{The poloidal $\langle B^2_{\mathrm{pol}}\rangle$ (\textit{top}) and toroidal magnetic energy $\langle B^2_{\mathrm{tor}}\rangle$ (\textit{bottom}) over the whole range of differential rotation accessible from the simulations for the $\ellsum\ $-modes. The differential rotation increases to the \textit{right} for each flux emergence rate, which is indicated by the background shade. The background shade becomes darker with increasing flux emergence rate. The lower DR simulations are displayed as diamonds and the higher DR simulations as circles and the simulated Sun is marked by the solar symbol $\odot$. The same symbol colour scheme is used as in Fig.~\ref{fig:BtorBpolRepeatori}.}
  \label{fig:TP_EtotvsStars_lowDR}
\end{figure}

In Figure~\ref{fig:TP_EtotvsStars_lowDR} we plot the magnetic energy of the poloidal $\langle B^2_{\mathrm{pol}}\rangle$ (\textit{top}) and toroidal $\langle B^2_{\mathrm{tor}}\rangle$ (\textit{bottom}) component logarithmically for the $\ellsum\ $-modes over the differential rotation rates. The background shade indicates the different flux emergence rates of the simulations and gets darker with increasing flux emergence rate. For each flux emergence rate the differential rotation increases to the right. We plot the entire range of the differential rotation considered with the lower and higher DR simulations starting from $0.3$\,\DRSun\ \ to $5$\,\DRSun\ . The lower DR simulations are plotted as diamonds and the higher DR simulations as circles. The errors are similar or smaller than the plot symbols. 

The poloidal and toroidal energy increase with flux emergence rate for all $\ellsum\ $-modes and differential rotation rates, see Fig.~\ref{fig:TP_EtotvsStars_lowDR}. 
Figure~\ref{fig:TP_EtotvsStars_lowDR} (\textit{top}) displays a widely linear decrease for $\log(\langle B^2_{\mathrm{pol}}\rangle)$ with differential rotation. We see this trend for all flux emergence rates but it is strongest for the low flux emergence rates. 
Figure~\ref{fig:TP_EtotvsStars_lowDR} (\textit{bottom}) shows a weak decrease in $\langle B^2_{\mathrm{tor}}\rangle$ for the higher $\ellsum\ $-modes. However, the lower $\ellsum\ $-modes remain widely constant or present an increase with differential rotation, especially for the solar flux emergence rates. $\langle B^2_{\mathrm{tor}}\rangle$ increases by approximately one magnitude from the dipolar mode $\ell = 1$ to the cumulative quadrupolar mode $\ellsum\  \le 2$. This increase appears stronger for higher flux emergence rates.

\begin{figure} 
	\includegraphics[width=\columnwidth]{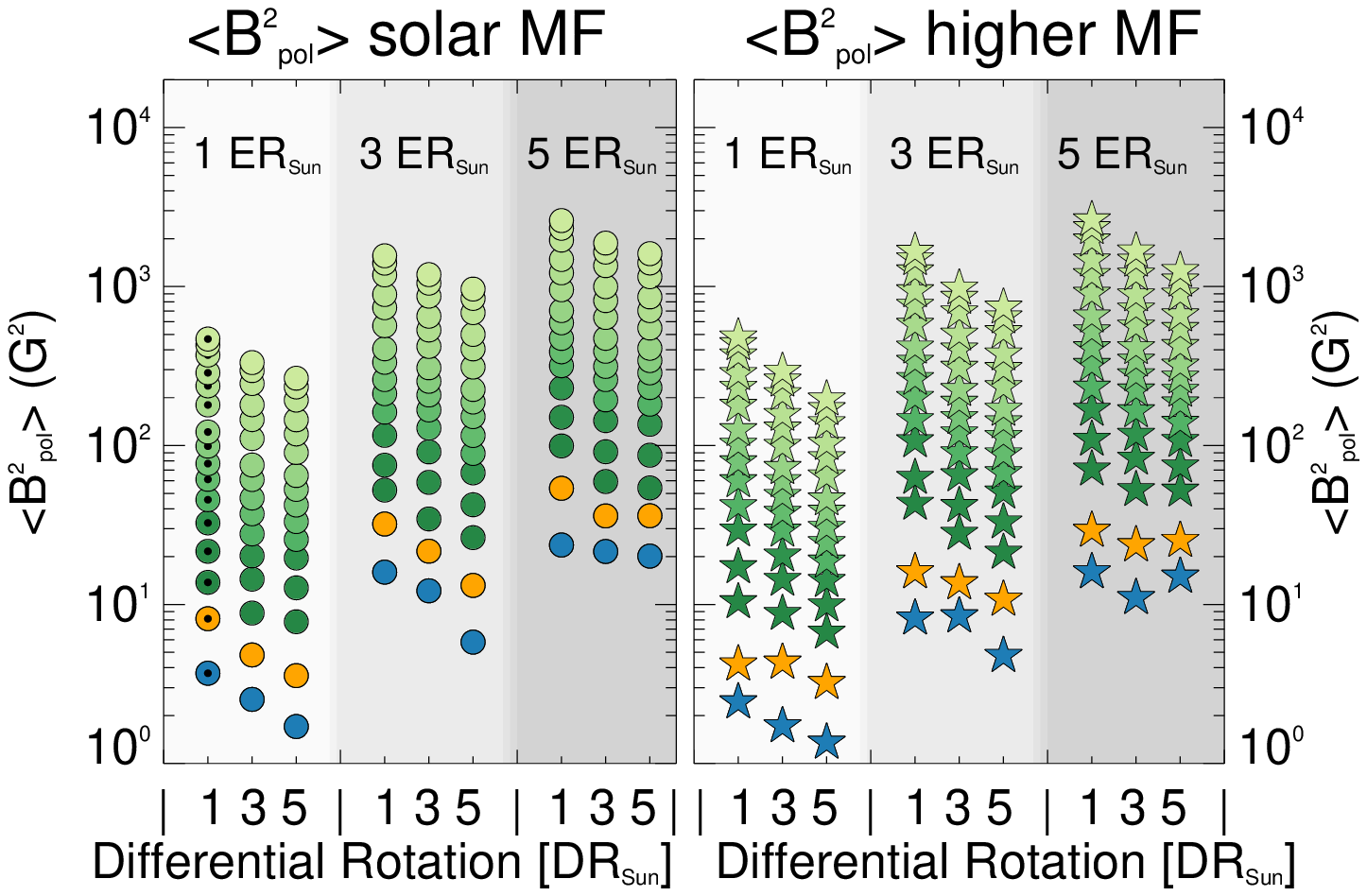}
	\includegraphics[width=\columnwidth]{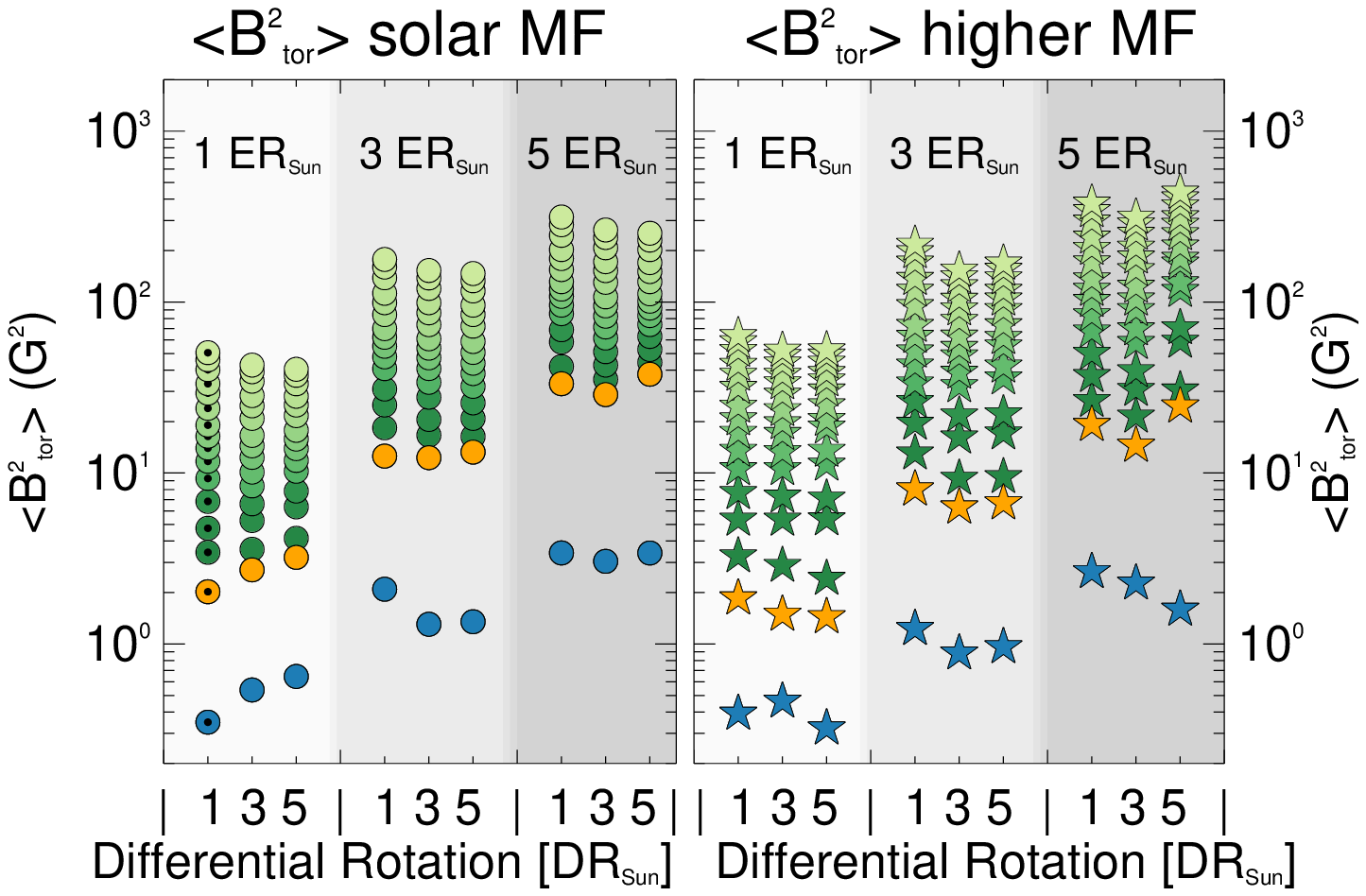}
   \caption{The poloidal $\langle B^2_{\mathrm{pol}}\rangle$ (\textit{top}) and toroidal magnetic energy $\langle B^2_{\mathrm{tor}}\rangle$ (\textit{bottom}) over the higher differential rotation range for the $\ellsum\ $-modes comparing the solar meridional flow simulations (\textit{left}, circles) with the higher meridional flow simulations (\textit{right}, stars). Beside that the same format as in Fig.~\ref{fig:TP_EtotvsStars_lowDR} is used. }
  \label{fig:TP_EtotvsStars_highMF}
\end{figure}

Figure~\ref{fig:TP_EtotvsStars_highMF} compares the simulations of solar and increased meridional flow, for the higher DR range ($1-5$\,\DRSun\ ) again for the poloidal $\langle B^2_{\mathrm{pol}}\rangle$ (\textit{top}) and the toroidal energy $\langle B^2_{\mathrm{tor}}\rangle$ (\textit{bottom}). A similar format to Fig.~\ref{fig:TP_EtotvsStars_lowDR} is used. The results for the solar MF (circle symbol, \textit{left} panel) and the higher MF simulations (star symbol, \textit{right} panel) are displayed next to each other. The background shade indicates again the flux emergence rate and the differential rotation increases to the right from $1$\,\DRSun\ \ to $5$\,\DRSun\ .

The higher MF has only a small effect on the poloidal energy $\langle B^2_{\mathrm{pol}}\rangle$, see Fig.~\ref{fig:TP_EtotvsStars_highMF} (\textit{top} panel). Only the lowest $\ellsum\ $-modes seem to have slightly weaker $\langle B^2_{\mathrm{pol}}\rangle$. However, the toroidal energy $\langle B^2_{\mathrm{tor}}\rangle$ changes with the higher MF, see Fig.~\ref{fig:TP_EtotvsStars_highMF} \textit{bottom}. In general, $\langle B^2_{\mathrm{tor}}\rangle$ becomes stronger for the high $\ellsum\ $-modes, but weaker for the low $\ellsum\ $-modes. The even low $\ellsum\ $-modes rise more strongly than the odd low $\ellsum\ $-modes and the gap between the dipolar and cumulative quadrupolar mode is smaller than for the solar MF.

The analysis of the quadrupole/dipole $\tfrac{\langle B^2_{l=2}\rangle}{\langle B^2_{l=1}\rangle}$ and octopole/dipole ratio $\tfrac{\langle B^2_{l=3}\rangle}{\langle B^2_{l=1}\rangle}$ of the total magnetic energy shows no clear dependencies on the differential rotation and meridional flow, (not shown). The octopole/dipole ratio is larger than the quadrupole/dipole ratio for the solar flux emergence rate. For the simulated Sun the ratios are $\tfrac{\langle B^2_{l=2}\rangle}{\langle B^2_{l=1}\rangle} = 1.2$ and $\tfrac{\langle B^2_{l=3}\rangle}{\langle B^2_{l=1}\rangle} = 1.5$. For higher flux emergence rates this is reversed: the quadrupole/dipole ratio is now larger than the octopole/dipole ratio.

Summing up, the flux emergence rate increases the energy in both components $\langle B^2_{\mathrm{pol}}\rangle$ and $\langle B^2_{\mathrm{tor}}\rangle$ and for all $\ellsum\ $-modes. A higher differential rotation causes a decrease in $\langle B^2_{\mathrm{pol}}\rangle$ for all $\ellsum\ $-modes. $\langle B^2_{\mathrm{tor}}\rangle$ shows a weak decrease for the high $\ellsum\ $-modes and remains widely constant for the low $\ellsum\ $-modes. In general, $\langle B^2_{\mathrm{tor}}\rangle$ increases from the dipolar to the cumulative quadrupolar mode by one order of magnitude. A higher meridional flow results in a weak decrease of $\langle B^2_{\mathrm{pol}}\rangle$ for the low $\ellsum\ $-modes. $\langle B^2_{\mathrm{tor}}\rangle$ gets stronger for the high $\ellsum\ $-modes and weaker for the low $\ellsum\ $-modes. For the low even $\ellsum\ $-modes the rise in $\langle B^2_{\mathrm{tor}}\rangle$ is stronger than for the low odd $\ellsum\ $-modes.

\subsection{Distribution of magnetic energy across modes}
\label{subsec:Distribution}

\begin{figure*} 
	\includegraphics[width=2\columnwidth]{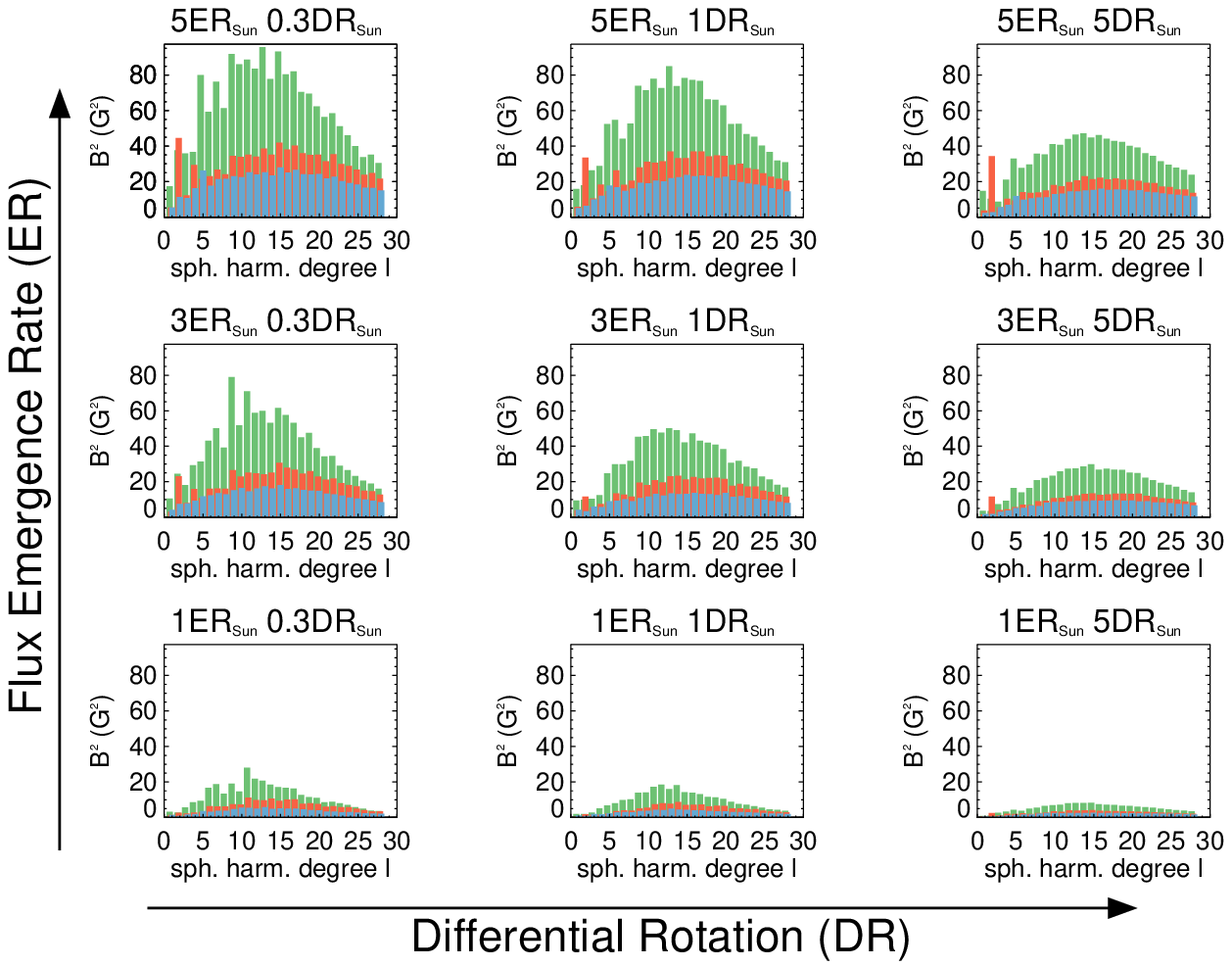}
  \caption{The magnetic energy distribution for the radial $\langle B^2_{\mathrm{rad}}\rangle(l)$ (green bars), azimuthal $\langle B^2_{\mathrm{azi}}\rangle(l)$ (red bars) and meridional component $\langle B^2_{\mathrm{mer}}\rangle(l)$ (blue bars). The title above each barplot indicates the flux emergence rate and differential rotation. The differential rotation increases horizontally and the flux emergence rate vertically. All here shown simulations have a solar meridional flow. For a comparison with the higher meridional flow simulations, see Figure~\ref{fig:BP_E_RadAziMerPolTor_MF} \textit{left column}.
	}
  \label{fig:BP_E_RadAziMer}
\end{figure*}

Figures~\ref{fig:TP_EtotvsStars_lowDR} and~\ref{fig:TP_EtotvsStars_highMF} show that the effect of differential rotation and flux emergence rate may be different for different modes. We therefore investigate the distribution of $\langle B^2\rangle(\ell)$ across all the modes.

Figure~\ref{fig:BP_E_RadAziMer} shows the energy distribution for the radial (green), azimuthal (red) and meridional component (blue) over the single $\ell$-modes ranging from $1-28$. The title of each barplot gives the flux emergence rate and the differential rotation of the simulation. The flux emergence rate increases in the vertical direction and the differential rotation in the horizontal direction. All simulations in Figure~\ref{fig:BP_E_RadAziMer} have a solar meridional flow. The influence of the higher meridional flow can be seen by comparing the solar MF simulations of the \textit{middle column} in Fig.~\ref{fig:BP_E_RadAziMer} with their corresponding higher MF simulations plotted at the \textit{left column} of Figure~\ref{fig:BP_E_RadAziMerPolTor_MF} in the appendix.

By analysing the individual values of the energy distributions we notice the following: The maximum energy per $\ell$-mode increases with flux emergence rate (vertical direction Fig.~\ref{fig:BP_E_RadAziMer}) and decreases with differential rotation (horizontal direction Fig.~\ref{fig:BP_E_RadAziMer}). The flux emergence rate has the greater effect. 

An increase in differential rotation in general produces a decrease of $\langle B^2_{\mathrm{rad}}\rangle$, $\langle B^2_{\mathrm{azi}}\rangle$ and $\langle B^2_{\mathrm{mer}}\rangle$ for most of the $\ell$-modes. This seems counter-intuitive to the prediction of the dynamo theory at first glance but we discuss this result in section~\ref{sec:Discussion} extensively. The only exception to the decrease are the azimuthal quadrupolar modes that remain roughly constant while the meridional and radial quadrupolar modes decrease for high differential rotation even faster than the dipolar modes. See also Figure~\ref{fig:BP_E_RadAziMer_zoom} in the appendix for a detailed view of the first ten $\ell$-modes.
The higher meridional flow causes an increase of $\langle B^2_{\mathrm{mer}}\rangle$ for the low $\ell$-modes at the cost of $\langle B^2_{\mathrm{rad}}\rangle$, see Fig.~\ref{fig:BP_E_RadAziMerPolTor_MF} \textit{left column}. The energy stored in the azimuthal quadrupolar mode is less for the higher MF than for the solar MF simulations.

The stellar properties influence the shape of the energy distributions. With increasing flux emergence rate the azimuthal quadrupolar modes become more and more dominant over the other azimuthal modes until the quadrupolar mode stores the highest $\langle B^2_{\mathrm{azi}}\rangle$. An increase in flux emergence rate enhances also the following trends with differential rotation. With increasing differential rotation the peak in the radial distribution becomes broader and the peak-value shifts slightly to higher $\ell$-modes. The radial component peaks in general at lower $\ell$-modes compared to the azimuthal and meridional component for the solar meridional flow. For higher MF, see Fig.~\ref{fig:BP_E_RadAziMerPolTor_MF} \textit{left column}, the maximum of the meridional component shifts to lower $\ell$-modes. The meridional component peaks therefore before the radial and azimuthal component. The azimuthal $\ell = 2$ mode becomes weaker with higher MF compared to the peak for the azimuthal component at mid $\ell$-modes. The peak distribution of the radial component becomes broader with increasing meridional flow.

\begin{figure*} 
	\includegraphics[width=2\columnwidth]{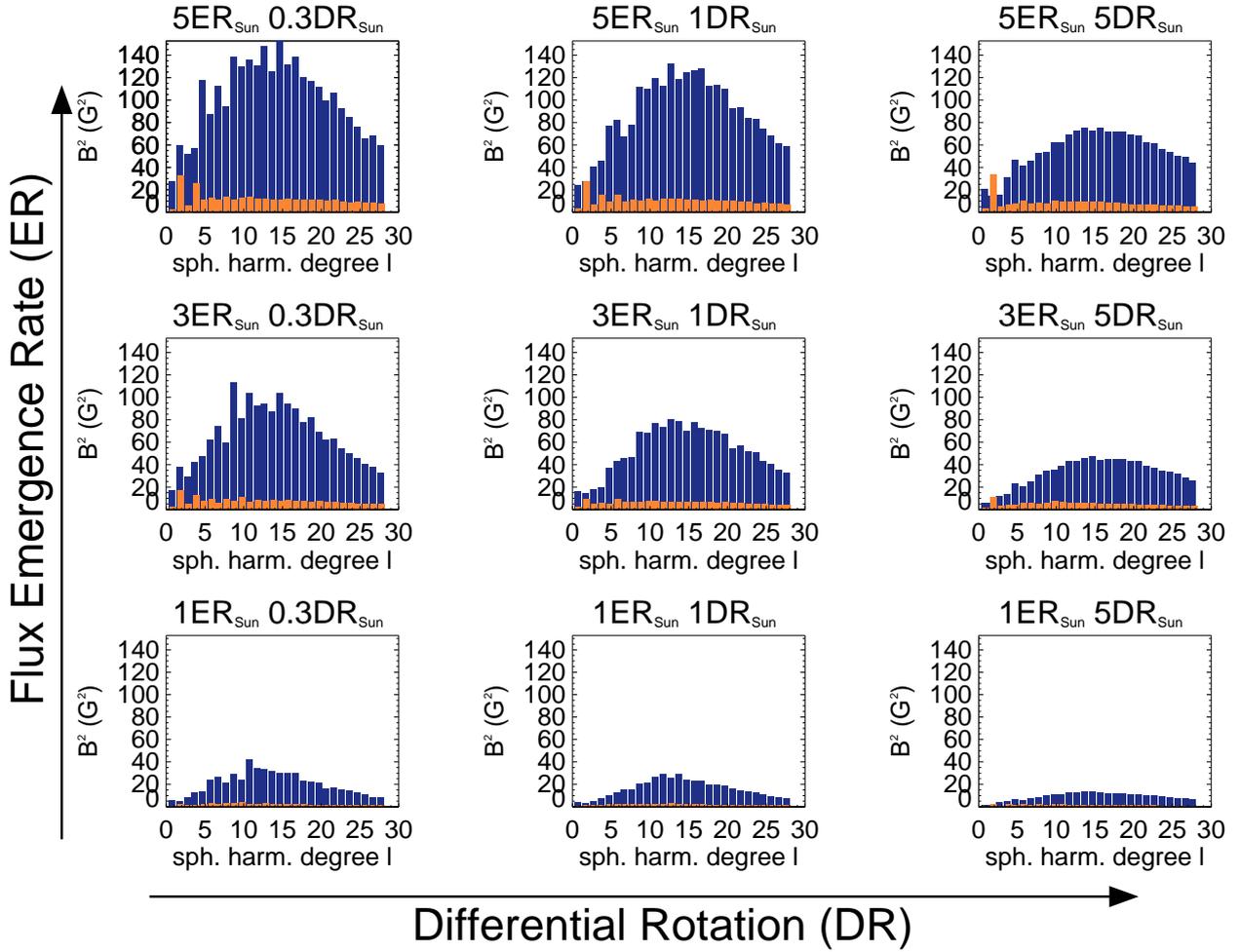}
  \caption{
	The magnetic energy distribution for the poloidal $\langle B^2_{\mathrm{pol}}\rangle(\ell)$ (plum bars) and toroidal component $\langle B^2_{\mathrm{tor}}\rangle(\ell)$ (orange bars). The same format as for Fig.~\ref{fig:BP_E_RadAziMer} is used. For a comparison with the higher meridional flow simulations, see Figure~\ref{fig:BP_E_RadAziMerPolTor_MF} \textit{right column}.
	}
  \label{fig:BP_E_TorPol}
\end{figure*}

A rather simpler picture emerges if we plot $\langle B^2_{\mathrm{pol}}\rangle(\ell)$ and $\langle B^2_{\mathrm{tor}}\rangle(\ell)$, see Fig.~\ref{fig:BP_E_TorPol}. 
The peak values for $\langle B^2_{\mathrm{pol}}\rangle(\ell)$ and $\langle B^2_{\mathrm{tor}}\rangle(\ell)$ increase again with flux emergence rate and decrease with differential rotation, whereby the effect of the flux emergence rate is the greatest. In general, we notice that the peak for the toroidal distribution (without the $\ell = 2$-mode) appears at lower $\ell$-modes than the peak for the poloidal distribution. 

Increasing the flux emergence rate increases the toroidal $\ell = 2$ mode relative to the peak of the toroidal component at mid $\ell$-modes, similar to the azimuthal component in Fig~\ref{fig:BP_E_RadAziMer}.
Increasing the  differential rotation decreases the magnetic energy in all $\ell$-modes of the poloidal component. The quadrupolar modes stand out again: the toroidal $\ell = 2$ mode is widely constant or increases slightly with increasing differential rotation while the poloidal $\ell = 2$ mode decreases even more compared to the other modes. For $\mrm{DR} \le 3\ $\DRSun \ \ the toroidal quadrupolar mode is the strongest toroidal mode. Additionally, the poloidal peak distribution becomes broader and shifts slightly to higher $\ell$-modes with increasing differential rotation.
Figure~\ref{fig:BP_E_RadAziMerPolTor_MF} \textit{right column} in the appendix provides a comparison of the solar MF with the higher MF simulations, see Fig.~\ref{fig:BP_E_TorPol} \textit{middle column} and Fig.~\ref{fig:BP_E_RadAziMerPolTor_MF} \textit{right column}.  Increasing the meridional flow enhances all low toroidal $\ell$-modes ($\ell \approx 2-16$) so that for the higher MF the strongest toroidal modes are always at mid $\ell$-modes. Especially the low even $\ell$-modes are enhanced for higher MF in connection with high flux emergence rate and differential rotation, but there is no preference to increase only the $\ell = 2$ mode, see Fig.~\ref{fig:BP_Ef_55MF} \textit{top row}. Next to this the poloidal peak distribution becomes broader with higher MF.

To summarise: the stellar properties affect the magnetic energy distributions. A higher flux emergence rate increases $\langle B^2\rangle$ for all components and $\ell$-modes. The azimuthal and toroidal quadrupolar $\ell = 2$ mode increases relative to the peak values of the azimuthal and toroidal distribution at mid $\ell$-modes and could become the strongest azimuthal or toroidal mode. A higher differential rotation decreases $\langle B^2\rangle$ for most components and $\ell$-modes. Exceptions are the azimuthal and toroidal quadrupolar $\ell = 2$ modes that remain widely constant. The radial, meridional and poloidal quadrupolar modes decrease faster than other modes with higher differential rotation. Next to this, the radial and poloidal peak distributions broaden and their peaks shift to higher $\ell$-modes. A higher meridional flow increases $\langle B^2_{\mathrm{mer}}\rangle$ at the cost of $\langle B^2_{\mathrm{rad}}\rangle$ and the peak of the $\langle B^2_{\mathrm{mer}}\rangle(\ell)$ distribution shifts to lower $\ell$-modes. The radial and poloidal peak distribution broadens. The lower toroidal $\ell$-modes are enhanced especially the even $\ell$-modes but the azimuthal and toroidal quadrupolar mode is weaker relative to the toroidal and azimuthal peak at mid $\ell$-modes compared to the solar meridional flow simulations.

\begin{figure*} 
	\includegraphics[width=2\columnwidth]{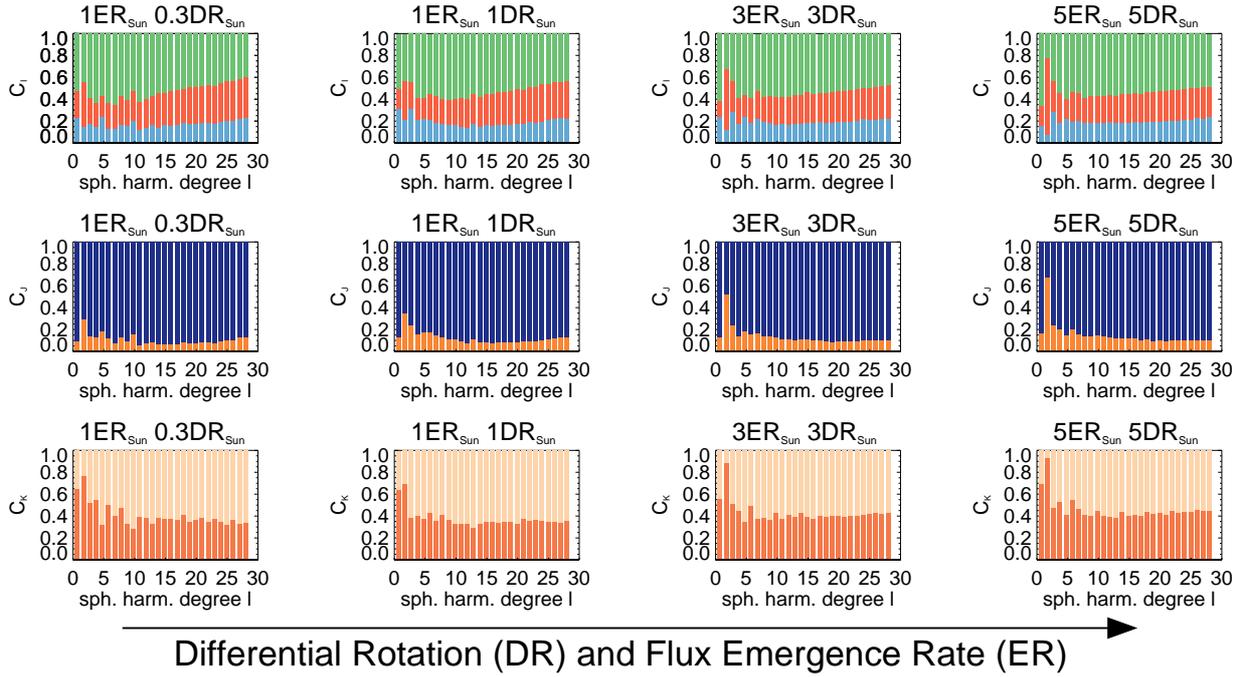}
    \caption{ The cumulative total $C(\ell) = \sum f(\ell)$ of the fractions for the different field components over $\ell$-modes for the solar meridional flow simulations. The \textit{top row} shows the cumulative total $C_I(\ell) = f_{\mrm{mer}}(\ell) + f_{\mrm{azi}}(\ell) + f_{\mrm{rad}}(\ell)$ of the meridional (blue), azimuthal (red) and radial (green) fraction. The \textit{middle row} the cumulative total $C_J(\ell) = f_{\mrm{tor}}(\ell) + f_{\mrm{pol}}(\ell)$ of the toroidal (orange) and poloidal (plum) fraction. The \textit{bottom row} displays the cumulative total $C_K(\ell) = f_{\mrm{azi,tor}}(\ell) + f_{\mrm{mer,tor}}(\ell)$ of the azimuthal toroidal (red orange) and the meridional toroidal (light orange) fraction, that build up the toroidal component. The title of each barplot indicates the flux emergence rate and differential rotation. Both parameters increase horizontally. For a comparison with the higher meridional flow simulations see Figure~\ref{fig:BP_f_All_MF}.
		}
    \label{fig:BP_f_All}
\end{figure*}

\begin{figure} 
	\includegraphics[width=\columnwidth]{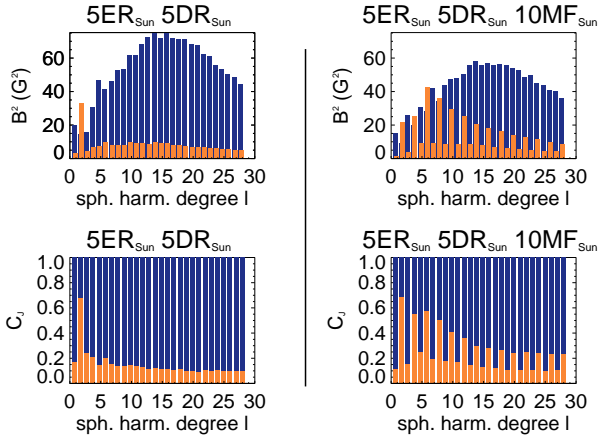}
    \caption{The simulations of an active star ($\mrm{DR} = 5$\,\DRSun\ \ and $\mrm{ER} = 5$\,\ERSun\ ) with solar meridional flow, \textit{left column}, and higher meridional flow $\mrm{MF} = 10$\,\MFSun\ , \textit{right column}. The \textit{top row} displays the magnetic energy distribution for the poloidal $\langle B^2_{\mathrm{pol}}\rangle(\ell)$ (plum bars) and toroidal component $\langle B^2_{\mathrm{tor}}\rangle(\ell)$ (orange bars) and the \textit{bottom row} the cumulative total $C_J(\ell) = f_{\mrm{tor}}(\ell) + f_{\mrm{pol}}(\ell)$ of the toroidal (orange) and poloidal (plum) fraction. 
		}
    \label{fig:BP_Ef_55MF}
\end{figure}

Figures~\ref{fig:BP_E_RadAziMer} and \ref{fig:BP_E_TorPol} suggest that the response of the $\ell = 2$ mode to changes in flux emergence rate and differential rotation is different to the response of the other modes. This is even more clear by analysing the fractions.
The fractions of the different field components $f(\ell)$ are displayed in Figure~\ref{fig:BP_f_All}.
On the \textit{top row} of Fig.~\ref{fig:BP_f_All} we plot the cumulative total $C_{I}(\ell)$ of the radial, azimuthal and meridional fraction of the total energy distribution per $\ell$-mode.
\begin{align}
C_I(\ell) &= \sum_i f_{i}(\ell) \\
f_{i}(\ell) &= \frac{\langle B^2_{\mathrm{i}}\rangle(\ell)}{\langle B^2_{\mathrm{tot}}\rangle(\ell)}, i \in \mrm{(rad, azi, mer)} \\
\langle B^2_{\mathrm{tot}}\rangle(\ell) &= \langle B^2_{\mathrm{rad}}\rangle(\ell)+\langle B^2_{\mathrm{azi}}\rangle(\ell)+\langle B^2_{\mathrm{mer}}\rangle(\ell)
\end{align}
The radial component is again green, the azimuthal component red and the meridional component blue. The second row of Fig.~\ref{fig:BP_f_All} displays the cumulative total $C_{J}(\ell)$ of the poloidal (plum) and toroidal fraction (orange).
\begin{align}
C_J(\ell) &= \sum_j f_{j}(\ell) \\
f_{j}(\ell) &= \frac{\langle B^2_{\mathrm{j}}\rangle(\ell)}{\langle B^2_{\mathrm{tot}}\rangle(\ell)}, j \in \mrm{(pol, tor)} \\
\langle B^2_{\mathrm{tot}}\rangle(\ell) &= \langle B^2_{\mathrm{pol}}\rangle(\ell)+\langle B^2_{\mathrm{tor}}\rangle(\ell)
\end{align}
The \textit{bottom row} of Fig.~\ref{fig:BP_f_All} presents the cumulative total $C_{K}(\ell)$ of the azimuthal toroidal and meridional toroidal fraction of the toroidal energy $\langle B^2_{\mathrm{tor}}\rangle$.
\begin{align}
C_K(\ell) &= \sum_k f_{k}(\ell) \\
f_{k}(\ell) &= \frac{\langle B^2_{\mathrm{k}}\rangle(\ell)}{\langle B^2_{\mathrm{tor}}\rangle(\ell)}, k \in \mrm{(azi,tor; mer,tor)} \\
\langle B^2_{\mathrm{tor}}\rangle(\ell) &= \langle B^2_{\mathrm{azi,tor}}\rangle(\ell)+\langle B^2_{\mathrm{mer,tor}}\rangle(\ell)
\end{align}
The azimuthal toroidal fraction is coloured in red orange and the meridional toroidal fraction in light orange.
We show the results for the solar meridional flow simulations with increasing flux emergence rate and differential rotation. For the higher meridional flow simulation see Figure~\ref{fig:BP_f_All_MF} in the appendix.

The fractions of the radial $f_{\mrm{rad}}$, azimuthal $f_{\mrm{azi}}$ and meridional component $f_{\mrm{mer}}$ are widely unchanged for the high $\ell$-modes, see Fig.~\ref{fig:BP_f_All} \textit{top row}. There is only a small increase in $f_{\mrm{rad}}$ while $f_{\mrm{azi}}$ and $f_{\mrm{mer}}$ slightly decrease.  The radial component has the largest fraction with $f_{\mrm{rad}} \approx 0.5$. The azimuthal and meridional components show similar fractions around $f_{\mrm{rad}} \approx f_{\mrm{azi}} \approx 0.25$ regardless of flux emergence rate or differential rotation. The stellar properties influence the lower $\ell$-modes and therefore the large-scale field much more strongly. With increasing differential rotation, the fraction of azimuthal field $f_{\mrm{azi}}$ increases in the quadrupolar $\ell = 2$ mode and for the highest differential rotation rates for some neighbouring $\ell$-modes, too. Next to this the radial dipolar $\ell = 1$ becomes mode more dominant. These effects are enhanced by an increasing flux emergence rate. Comparing the solar with higher meridional flow, see Fig.~\ref{fig:BP_f_All_MF} \textit{top}, we notice that the fraction of the meridional field $f_{\mrm{mer}}$ increases for the low $\ell$-modes mainly at the cost of the radial field for higher MF.

The fractions of the poloidal $f_{\mrm{pol}}$ and toroidal component $f_{\mrm{tor}}$ are as well mainly unchanged for the high $\ell$-modes with $f_{\mrm{pol}} \approx 0.9$ and $f_{\mrm{tor}} \approx 0.1$, see Fig.~\ref{fig:BP_f_All} \textit{middle row}. The toroidal fraction $f_{\mrm{tor}}$ increases with increasing differential rotation for the low $\ell$-modes, especially the quadrupolar $\ell = 2$ mode. This is due to the strong decrease of $\langle B^2_{\mathrm{pol}}\rangle$ while $\langle B^2_{\mathrm{tor}}\rangle$ remains constant. The dipolar $\ell = 1$ mode remains strongly poloidal, whereas the quadrupolar $\ell = 2$ mode becomes dominantly toroidal for $\mrm{DR} \ge 3$\,\DRSun\ . For the higher MF simulations, see Fig.~\ref{fig:BP_f_All_MF} \textit{middle}, the toroidal fraction $f_{\mrm{tor}}$ increases for the low $\ell$-modes. The even $\ell$-modes are especially enhanced. This is best seen for the more active simulated stars, see Fig.~\ref{fig:BP_Ef_55MF} \textit{bottom row}. The quadrupolar $\ell = 2$ mode is often stronger but not always the strongest toroidal mode in connection with higher MF. Simulations with high differential rotation and higher meridional flow show the highest $f_{\mrm{tor}}$ also in the $\ell = 4$ or $\ell = 6$ mode (not shown).

In order to better understand the behaviour of the toroidal field, we show the relative contributions of its azimuthal and meridional parts. The lowest $\ell$-modes are highly azimuthal toroidal (red orange bars), whereas the higher $\ell$-modes show roughly equal fractions for the meridional toroidal $f_{\mrm{mer,tor}}$ (light orange bars) and azimuthal toroidal field $f_{\mrm{azi,tor}}$. The quadrupolar $\ell = 2$ mode has the highest $f_{\mrm{azi,tor}}$ for the simulations with solar meridional flow. We see an increase in $f_{\mrm{azi,tor}}$ with increasing differential rotation for both the quadrupolar $\ell = 2$ mode and the highest $\ell$-modes. These effects are again enhanced by an increasing flux emergence rate. With increasing meridional flow, see Fig.~\ref{fig:BP_f_All_MF} \textit{bottom}, the meridional toroidal fraction $f_{\mrm{mer,tor}}$ also increases for the lowest $\ell$-modes and the azimuthal toroidal fraction $f_{\mrm{azi,tor}}$ for the highest $\ell$-modes. Even the strong azimuthal toroidal $\ell = 2$ mode can be suppressed by this effect of increased meridional toroidal field for the low $\ell$-modes, see Fig.~\ref{fig:BP_f_All_MF} \textit{bottom}. However, with stronger differential rotation the $\ell = 2$ and further even $\ell$-modes become dominant azimuthal toroidal again (not shown).

Furthermore, we investigate the average of the large-scale field ($\ell = 1-5$) and the average over all $\ell$-modes for the toroidal and azimuthal fraction, (not shown). Also the average over the large-scale field and over all $\ell$-mode confirms that the fraction of toroidal field $f_{\mrm{tor}}$ increases with differential rotation and that the flux emergence rate enhances this effect. The toroidal fraction $f_{\mrm{tor}}$ increases due to a decrease in $\langle B^2_{\mathrm{pol}}\rangle$, while $\langle B^2_{\mathrm{tor}}\rangle$ remains widely constant. These trends are also noticeable for the large-scale field average for azimuthal fraction $f_{\mrm{azi}}$ but not for the average over all $\ell$-modes. With increasing differential rotation $f_{\mrm{tor}}$  doubles for the averaged large-scale field from $\approx 0.15$ to $\approx 0.3$. Even when we average over all the $\ell$-modes including the large and small-scale field we see a small increase of $0.05$ in $f_{\mrm{tor}}$. The increase of the large-scale field average for $f_{\mrm{azi}}$ is smaller from $\approx 0.25$ to $\approx 0.35$. It shows that the differential rotation is changing the field topology to more toroidal and slightly more azimuthal field topologies. 

To sum up: the flux emergence rate enhances the following effects of the differential rotation on the fractions of the different field components. The fraction of the toroidal $f_{\mrm{tor}}$ and azimuthal field $f_{\mrm{azi}}$ increases for the average of the lower $\ell$-modes representing the large-scale field. The azimuthal and toroidal quadrupolar modes rise the most. Also the azimuthal toroidal fraction $f_{\mrm{azi,tor}}$ increases for the quadrupolar mode as well as for the higher $\ell$-modes. A higher meridional flow increases  $f_{\mrm{mer}}$ by decreasing $f_{\mrm{rad}}$ for the low $\ell$-modes. The toroidal fraction $f_{\mrm{tor}}$ rises especially for the even $\ell$-modes. For the low $\ell$-modes the fraction of the meridional toroidal field $f_{\mrm{mer,tor}}$ increases while for the high $\ell$-modes the fraction of the azimuthal toroidal field $f_{\mrm{azi,tor}}$ rises with higher meridional flow.

\subsection{The axisymmetry of different modes}
\label{subsec:Axisymmetry}

\begin{figure} 
	\includegraphics[width=\columnwidth]{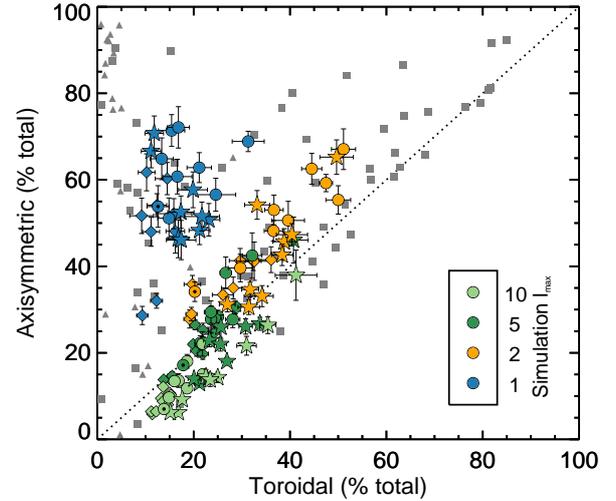}
  \caption{The percentages of the axisymmetric and toroidal fields for the simulations (coloured symbols) and for the observations (grey symbols). The figure presents the cumulative $\ellsum\ $-modes $\ellsum\  \le \ell_{\mrm{max}}, \ell_{\mrm{max}} = 1,2,5,10$, for the simulations. The lower DR simulations are plotted as diamonds, the higher DR simulations as circles and the higher MF simulations as stars. The same format as in Fig.~\ref{fig:BtorBpolRepeatori} is used and this figure is comparable to \citet{Lehmann2017}, Fig.~3.}
  \label{fig:AxiTor}
\end{figure}

In this subsection we investigate the axisymmetry of the surface magnetic field of the simulations. We do this first with respect to the fraction of the toroidal field $f_{\mrm{tor}}$, motivated by the results from \cite{See2015} that strongly toroidal fields are also strongly axisymmetric fields for observed stars. We define the axisymmetric fraction by dividing the magnetic energy stored in the axisymmetric mode ($m=0, \ellsum\ $) by the energy stored in all $m$-modes ($ \sum m, \ellsum\ $) per $\ellsum\ $-mode.
\begin{align}
f_{\mrm{axi},\ellsum\ } &= \frac{\langle B^2_{m=0, \ellsum\ }\rangle}{\sum\limits_m \langle B^2_{m, \ellsum\ }\rangle}
\end{align}

Figure~\ref{fig:AxiTor} shows the axisymmetric percentage over the toroidal percentage. The same format as for Fig.~\ref{fig:BtorBpolRepeatori} is used. The results for the optimal averaged simulations are plotted as diamonds for the lower DR, as circles for the higher DR and as stars for higher MF simulations. The errorbars are attached to each symbol and are not visible if the error is smaller than the plot symbol.

We see a clear trend with the $\ellsum\ $-modes in Fig.~\ref{fig:AxiTor}. That is consistent with \cite{Lehmann2017}. The dipolar modes (blue) are mainly poloidal and axisymmetric. The cumulative quadrupolar modes (orange) are strongly toroidal and axisymmetric. With increasing $\ellsum\ $-modes (greenish symbols) both the toroidal and the axisymmetric fraction decrease. The errors become generally smaller with increasing $\ellsum\ $-modes. For a direct comparison between the simulations and the observations focus on the greenish symbols for the $\ellsum\  \le 5, 10$ modes. The simulations cover a subsection of the observations. 
There are only very weak trends with differential rotation and meridional flow: the lower DR simulations own smaller fractions of toroidal and axisymmetric fields, whereas the higher MF simulations have higher fractions of toroidal field but similar fractions of axisymmetric fields compared to the higher DR simulations. 

\begin{figure} 
 	\includegraphics[width=\columnwidth]{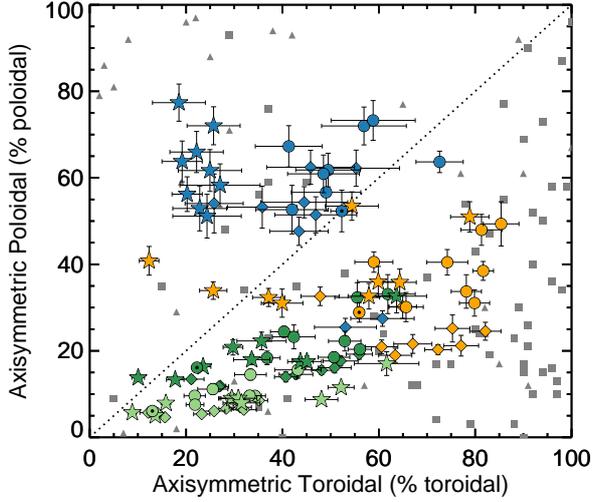}
  \caption{The percentages of the axisymmetric poloidal field and the axisymmetric toroidal field for the simulations. Same format as Fig.~\ref{fig:AxiTor}.}
  \label{fig:PolAxiTorAxi}
\end{figure}

The main contributors to the axisymmetry of the field can be seen in Figure~\ref{fig:PolAxiTorAxi}, motivated by Fig.~7, \cite{See2015}.
The axisymmetry of the dipolar modes (blue) is mainly due to the poloidal field, whereby the axisymmetry of the quadrupolar (orange) and of the other low $\ellsum\ $-modes (dark green) is mainly due to the toroidal field. Moving to the high $\ellsum\ $-modes the axisymmetry becomes very low. The errors for the axisymmetric toroidal field are often larger than for the axisymmetric poloidal field as the fraction of poloidal field is larger than the fraction of toroidal field for the majority of the $\ellsum\ $-modes. An increasing differential rotation increases $f_{\mrm{axi,pol}}$ and $f_{\mrm{axi,tor}}$ for all $\ellsum\ $-modes beside the dipolar modes. For the higher MF simulations the dipolar modes strikingly show a lack of axisymmetric toroidal field while owning the same percentage of axisymmetric poloidal field as the solar MF simulations. 

\begin{figure} 
	\includegraphics[width=\columnwidth]{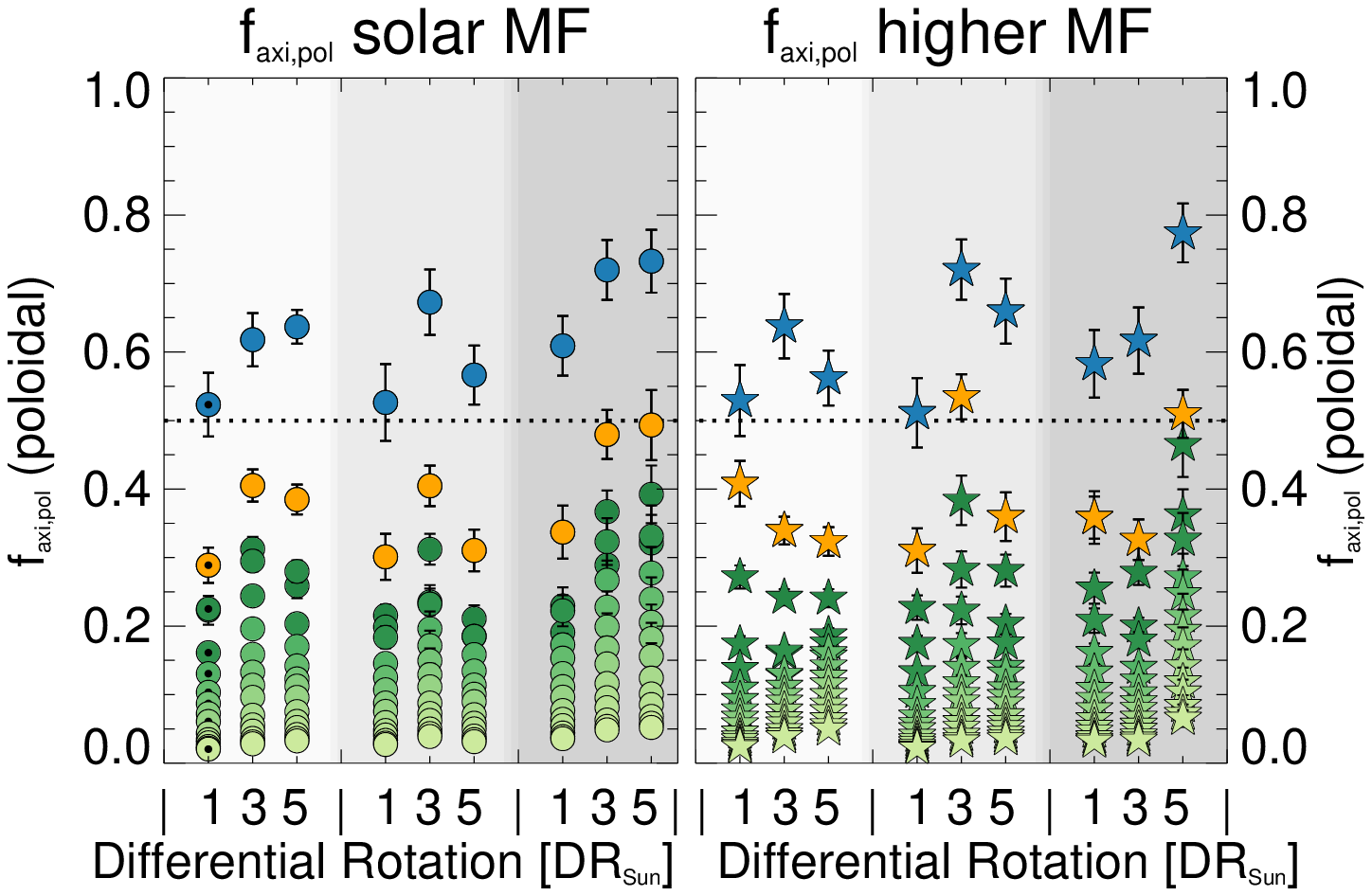}
	\includegraphics[width=\columnwidth]{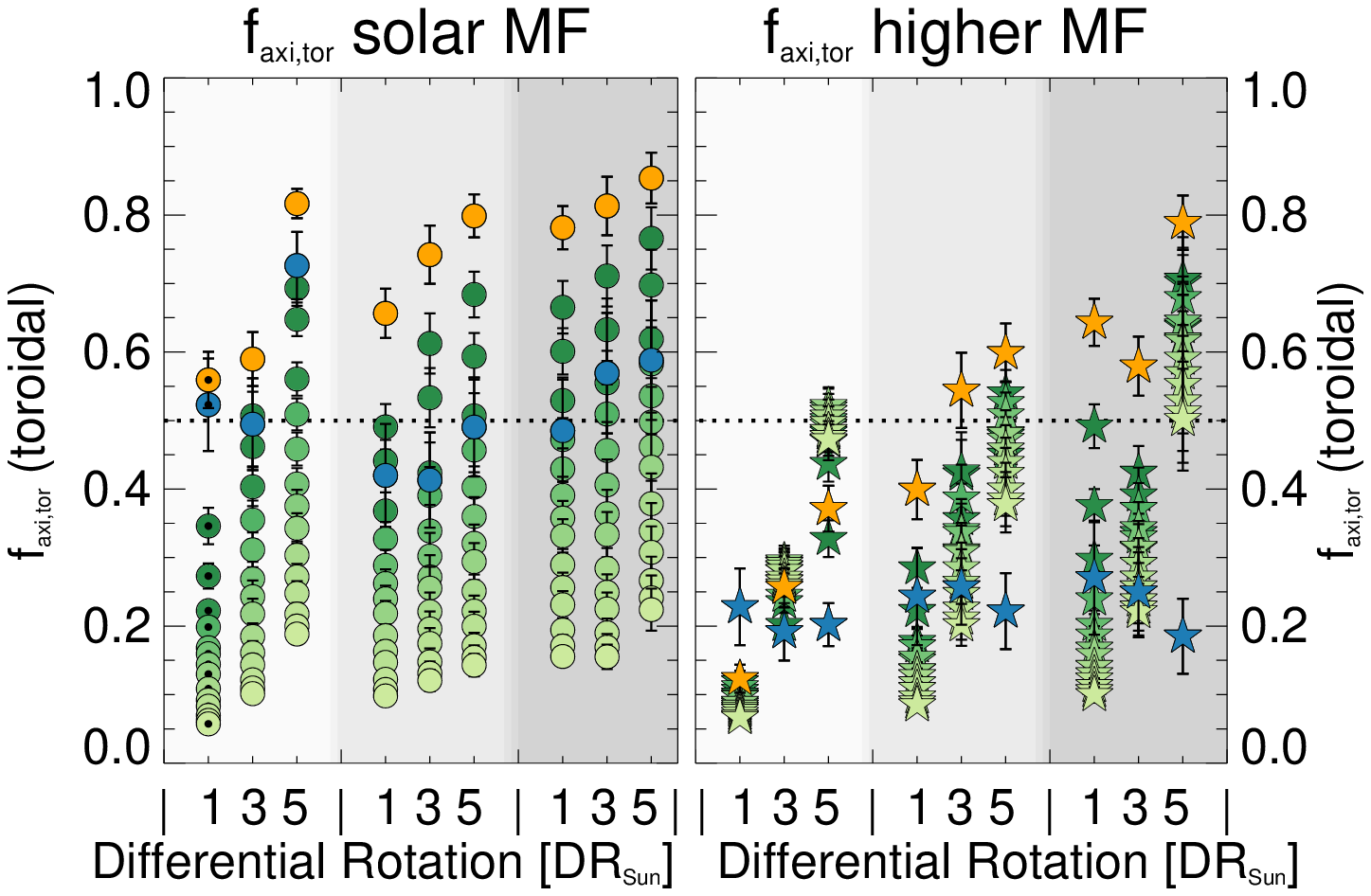}
    \caption{The fractions of the axisymmetric poloidal $f_{\mrm{axi,pol}}$ field (\textit{top}) and of the axisymmetric toroidal $f_{\mrm{axi,tor}}$ field (\textit{bottom}) for the cumulative $\ellsum\ $-modes comparing the solar meridional flow simulations (\textit{left}) with the higher meridional flow simulations (\textit{right}). The same format as in Fig.~\ref{fig:TP_EtotvsStars_highMF} is used.}
    \label{fig:TP_FaxipolaxitorvsStars_highMF}
\end{figure}

To further investigate the behaviour of the axisymmetric fields we plot the fraction of the axisymmetric poloidal $f_{\mrm{axi,pol}}$ and axisymmetric toroidal field $f_{\mrm{axi,tor}}$ for the solar and higher MF simulations, see Fig.~\ref{fig:TP_FaxipolaxitorvsStars_highMF}. A similar format to Fig.~\ref{fig:TP_EtotvsStars_highMF} is used. The \textit{top} panel shows the axisymmetric poloidal fraction $f_{\mrm{axi,pol}}$ and the \textit{bottom} panel the axisymmetric toroidal fraction $f_{\mrm{axi,tor}}$ for the solar MF on the \textit{left} and for the higher MF simulations on the \textit{right}. The dashed line indicates a fraction of $0.5$, when the field component becomes dominant. The errorbars are attached to each plot symbol and are not seen if the error is smaller than the plot symbol. 
The axisymmetric poloidal fraction $f_{\mrm{axi,pol}}$ decreases with increasing $\ellsum\ $-modes, see Fig.~\ref{fig:TP_FaxipolaxitorvsStars_highMF} \textit{top}. The dipolar mode is by far the most axisymmetric poloidal mode for the simulations with solar MF and higher MF. In general there are no differences between the solar MF (\textit{left}) and higher MF (\textit{right}) simulations for $f_{\mrm{axi,pol}}$.

The cumulative quadrupolar $\ellsum\  \le 2$ mode for the solar MF simulations has the highest fraction of axisymmetric toroidal field $f_{\mrm{axi,tor}}$, see Fig.~\ref{fig:TP_FaxipolaxitorvsStars_highMF} \textit{bottom left}. With further increasing $\ellsum\ $-modes decreases $f_{\mrm{axi,tor}}$ similar to $f_{\mrm{axi,pol}}$ in the \textit{top} panel. The dipolar mode displays lower $f_{\mrm{axi,tor}}$ than the $\ellsum\  \le 2$ mode but for high flux emergence rates even lower values than the $\ellsum\  \le 4$ mode. However, $f_{\mrm{axi,tor}}$ is generally higher than $f_{\mrm{axi,pol}}$ beside the dipolar mode and we see an increase of $f_{\mrm{axi,tor}}$ for all $\ellsum\ $-modes higher than $\ellsum\  \le 2$ with differential rotation.
The $f_{\mrm{axi,tor}}$ for the higher MF behaves differently compared to the solar MF simulations, see Fig.~\ref{fig:TP_FaxipolaxitorvsStars_highMF} \textit{bottom right}. First of all $f_{\mrm{axi,tor}}$ is generally lower and the $\ellsum\ $-modes lie closer together. For the higher flux emergence rates the quadrupolar $\ellsum\  \le 2$ has the highest $f_{\mrm{axi,tor}}$ and  $f_{\mrm{axi,tor}}$ decreases with increasing $\ellsum\ $-modes. For the solar flux emergence rate we see different, non-consistent and nearly an opposite trend with $\ellsum\ $-modes. Furthermore, the dipolar modes show $f_{\mrm{axi,tor}} \approx 0.2$ for all higher MF simulations independent of differential rotation or flux emergence rate. This explains the clear offset for the dipolar modes to lower $f_{\mrm{axi,tor}}$ in Figure~\ref{fig:PolAxiTorAxi}.

Summarising: the higher the toroidal field fraction the higher the axisymmetry fraction. The dipolar and cumulative quadrupolar modes are the most axisymmetric modes. The axisymmetry of the dipolar modes is based on the axisymmetric poloidal field while the axisymmetry of the cumulative quadrupolar and of the other low $\ellsum\ $-modes is based on the toroidal axisymmetric field. A higher differential rotation enhances $f_{\mrm{axi,pol}}$ and $f_{\mrm{axi,tor}}$ for all $\ellsum\ $-modes higher than the dipolar mode although $f_{\mrm{axi,tor}} > f_{\mrm{axi,pol}}$ for all solar MF simulations. The higher meridional flow causes a lack of $f_{\mrm{axi,tor}}$ but shows no effect on the $f_{\mrm{axi,pol}}$ which is best seen for the dipolar modes.

\subsection{Summarising the large-scale magnetic field topology}

\begin{figure*} 
	\includegraphics[height=6.15cm, trim = {0 0 290 0}, clip]{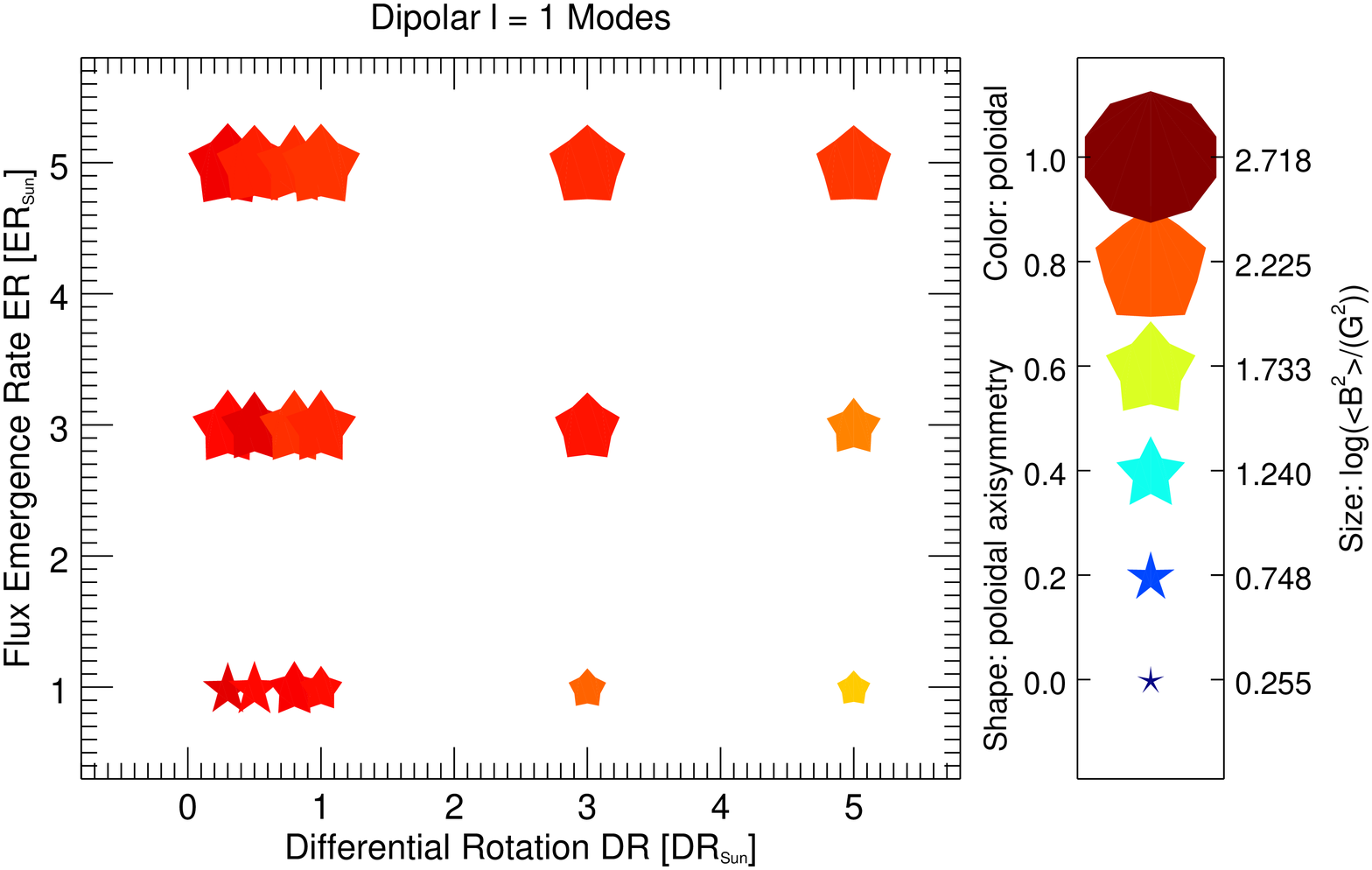}
	\includegraphics[height=6.15cm]{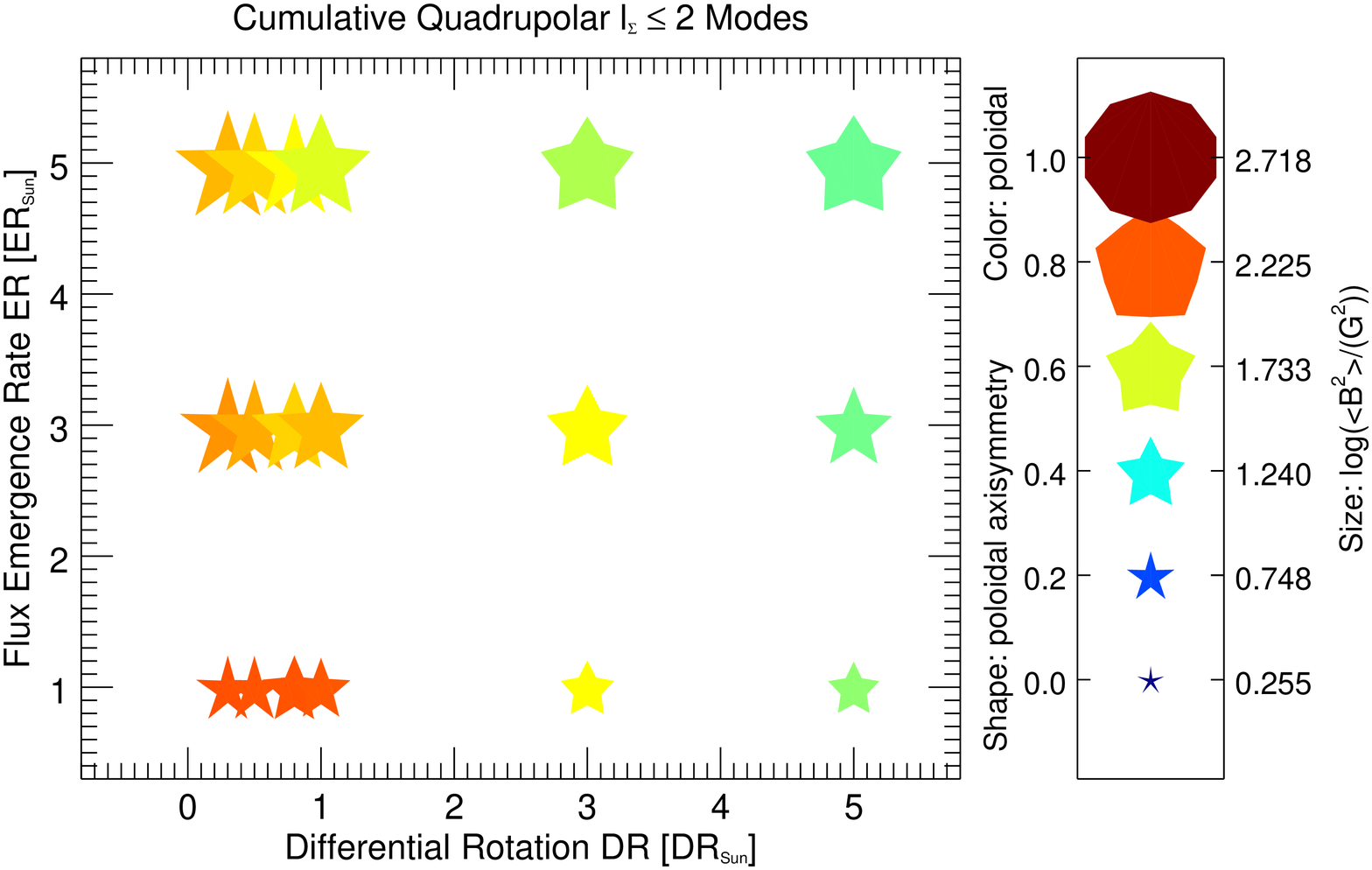}
    \caption{Summarising the properties of the dipolar $\ell = 1$ (\textit{left}) and cumulative quadrupolar $\ellsum\ \le 2$ mode (\textit{right}) for the solar meridional flow simulations. The higher meridional flow simulations are not shown. The flux emergence rate is plotted versus the differential rotation. The symbol size indicates the logarithmic magnetic energy $\log \langle B^2 \rangle$, the symbol shape the fraction of poloidal axisymmetric field $f_{\mrm{pol,axi}}$ and the symbol color the fraction of the poloidal field $f_{\mrm{pol}}$.}
    \label{fig:Confusogram}
\end{figure*}

Figure~\ref{fig:Confusogram} summarises some of the basic properties of the dipolar $\ell = 1$ (\textit{left}) and cumulative quadrupolar $\ellsum\ \le 2$ (\textit{right}) magnetic field topology in the same format as often use for the observed stellar magnetic field topologies, e.g. see \cite{Donati2008, Morin2008a, Morin2010, Jeffers2014, Folsom2016}. A direct comparison with the observational results is not adequate as Fig.~\ref{fig:Confusogram} shows only the $\ell = 1$ and $\ellsum\ \le 2$ mode. The observed magnetic field topologies include various numbers of $\ellsum\ $-modes depending on the stellar properties and observational constraints. However, Fig.~\ref{fig:Confusogram} provides an easier comparison between the topology of the main modes of the large-scale field of the simulations and the observed large-scale field topologies. The symbol size indicates the logarithmic magnetic energy $\log(\langle B^2 \rangle )$, which increases with flux emergence rate as seen for both modes and decreases slightly with differential rotation as best seen for the $\ell = 1$ modes. The symbol shape relates to the fraction of poloidal axisymmetric field $f_{\mrm{pol,axi}}$. The $\ell = 1$ modes show in general a higher $f_{\mrm{pol,axi}}$ than the $\ellsum\ \le 2$ modes but there are no clear trends of $f_{\mrm{pol,axi}}$ with differential rotation or flux emergence rate. The symbol color indicates the fraction of poloidal field $f_{\mrm{pol}}$. All the simulations show a poloidal dominated dipolar mode but the $\ellsum\ \le 2$ modes show a much lower fraction of poloidal field and therefore a higher fraction of toroidal field as $f_{\mrm{tor}} = 1-f_{\mrm{pol}}$. The increase of the toroidal field fraction $f_{\mrm{tor}}$ with differential rotation is best seen for the $\ellsum\ \le 2$ modes. The toroidal fraction $f_{\mrm{tor}}$ increases with flux emergence rate as well, but not as strongly as with differential rotation. These trends with differential rotation and flux emergence rate are still visible by including higher $\ellsum\ $-modes, whereby the axisymmetry of the poloidal field  $f_{\mrm{pol,axi}}$ as well as the fraction of toroidal field $f_{\mrm{tor}}$ decreases by including higher $\ellsum\ $-modes.

\section{Discussion}
\label{sec:Discussion}

By analysing the large- and the small-scale magnetic field topology of several non-potential flux transport simulations we want to understand the observed magnetic field topologies that are restricted to the large-scale field. We are especially interested in the influence of the small-scale field on the large-scale field topology and how different stellar parameters influence the magnetic field topology. We find that the large-scale field of the simulated solar-based stars fits the observed solar-like stars.
The simulated stars vary in flux emergence rate, differential rotation and meridional flow and their magnetic field topology, especially their large-scale magnetic field topology, depends on these stellar properties.
We conclude that the large- and small-scale magnetic field topology is determined by the global dipolar field and the small-scale flux pattern at the surface. The flux pattern results from the surface flux transport processes and the underlying flux emergence pattern. 

The most important spherical harmonic modes for the large-scale field are the poloidal dominated dipolar $\ell = 1$ modes and the quadrupolar $\ell = 2$ modes, which often show the highest fraction of toroidal field. The global dipole field of the stars is responsible for the dipolar modes. They are poloidal dominated and highly poloidal axi\-symmetric. The toroidal quadrupolar modes result from the large-scale properties of the solar flux emergence pattern: the bipoles appear at mid to low latitudes with opposite polarities on the different hemispheres. This pattern is best represented by the toroidal quadrupolar mode of the spherical harmonics for the large-scale field and is toroidal axisymmetric. Also the increase of one order of magnitude for the toroidal field from the dipolar to the cumulative quadrupolar mode follows from that. If we would have emerged a different pattern, e.g. bipoles at the equator with the same polarity, another spherical harmonic mode of the large-scale field would capture this pattern, e.g. in this case the toroidal dipolar mode. The lower spherical harmonic modes are sensitive to the large-scale properties of the small-scale flux, i.e. the averaged position and the polarity pattern of the bipoles. The higher spherical harmonic modes are sensitive to the size and shape of the bipoles. These modes are less axisymmetric and show a widely fixed ratio between the field components (toroidal/poloidal and radial/azimuthal/meridional) regardless of the stellar parameters. The bipoles show a larger area of radial field than of the azimuthal and meridional field, so that the radial energy distribution $\langle B^2_{\mathrm{rad}}\rangle(\ell)$ peaks at lower $\ell$-modes than the azimuthal $\langle B^2_{\mathrm{azi}}\rangle(\ell)$ and the meridional $\langle B^2_{\mathrm{mer}}\rangle(\ell)$ one.

Furthermore, the striking $\ell = 2$ quadrupolar mode could be a specific signature of a certain solar cycle phase. The flux emergence pattern for simulations is based on the average properties of the solar flux emergence pattern between January 2000 and January 2001. The Sun was at activity maximum of the activity cycle~23 at that time. The dipolar field was still detectable and the sunspots appeared mainly at mid-latitudes. Using the flux emergence pattern at the activity minimum or at the increasing or decreasing phase would probably change the behaviour of the dipolar and quadrupolar modes and the large-scale magnetic field topology.

The small-scale magnetic field topology is captured by the higher $\ellsum\ $-modes, which show a powerlaw behaviour, $\langle B^2_{\mathrm{tor}}\rangle \propto \langle B^2_{\mathrm{pol}}\rangle^{0.77\pm0.02}$, that is remarkably similar to the powerlaw $\langle B^2_{\mathrm{tor}}\rangle \propto \langle B^2_{\mathrm{pol}}\rangle^{0.72\pm0.08}$ found for the low-mass M-dwarfs ($M_{\star} < 0.5\,\mrm{M_{\odot}}$) by \cite{See2015}. This could be a hint that the M-dwarfs are covered randomly by small-scale field that show a similar ratio between the poloidal and toroidal field to the emerging bipoles in our flux transport simulations. The key point is that the small-scale field must emerge randomly all over the surface and is not organised on a larger scale. The more massive stars follow a different powerlaw $\langle B^2_{\mathrm{tor}}\rangle \propto \langle B^2_{\mathrm{pol}}\rangle^{1.25\pm0.06}$ \citep{See2015}. The large-scale field of the flux transport simulation restricted to $\ellsum\ \le 5$ or $\ellsum\ \le 10$ seems to follow again this steeper powerlaw with increasing flux emergence rate. This could be a hint that the observed magnetic field topologies for the more massive stars are also influenced by flux emergence patterns that are organised on a large scale similar to our simulations. These conclusions are only based on the similar powerlaw behaviour. To drive more secure conclusions we need to extend the parameter range of the simulations to reproduce the more active and toroidal dominated stars and the low mass M-dwarfs. The simulations analysed here only recover the solar-like stars.

The stellar properties flux emergence rate, differential rotation and meridional flow affect the surface magnetic field topology.
We find that the flux emergence rate widely influences the surface magnetic field topology. The flux emergence rate indicates the time rate at which new bipoles and magnetic features emerge. The stellar surface is occupied by more magnetic features for a higher flux emergences rate. This causes the increase of the magnetic energy $\langle B^2\rangle$ for all field components as seen by \cite{Gibb2016}. The toroidal and azimuthal quadrupolar mode is enhanced relative to the mid and higher $\ell$-modes for an increasing flux emergence rate. The emerged bipoles follow the flux emergence pattern and emerge with opposite polarities on the different hemispheres at a relatively small latitude range, which is best captured by the increased toroidal quadrupolar mode. A higher flux emergence rate consolidates these large-scale properties of the flux emergence pattern. The mid and higher $\ell$-modes are sensitive to the length scale of the single bipoles themselves and increase moderately compared to the toroidal quadrupolar mode. The broadening of the peak distribution for the mid $\ell$-mode of the poloidal and radial energy distribution $\langle B^2\rangle(\ell)$ results from the higher probability of seeing bipoles of different size and shape when the number of emerged bipoles increases.

The differential rotation influences the magnetic field topology as well. The most dominant dependency is the decrease of the poloidal energy $\langle B^2_{\mrm{pol}}\rangle$ while the toroidal energy $\langle B^2_{\mrm{tor}}\rangle$ remains widely constant with increasing differential rotation. This causes a strong increase of the toroidal field fraction $f_{\mrm{tor}}$. With higher differential rotation the equator rotates faster than the poles. The bipoles become stretched and elongated. The polarity inversion line (PIL) increases in length and with it the area where the opposite polarities of the radial field cancel within the bipole. This causes the decrease we have discovered in the radial and poloidal field, while the toroidal and azimuthal toroidal field remains widely constant as there is no cancellation for these field components. We rather see a small increase in the toroidal energy $\langle B^2_{\mrm{tor}}\rangle$ for the high $\ell$-modes due to the presence of bipoles stretched in the toroidal direction. Our results agree with \cite{Gibb2016}, who found a more sheared and non-potential corona with increasing differential rotation.
Furthermore, our results are consistent with the study of \cite{Bonanno2016}, who presented dynamo simulations in connection with harmonic corona fields for solar-like stars. \cite{Bonanno2016} results reproduce our observed increase of the toroidal fraction for 3D non-potential flux transport simulations and the observed increase of the toroidal field fraction for low-mass fast-rotating stars, see \cite{See2015} and Section \ref{subsec:ConnectSimandObs}.
Next to this a higher differential rotation causes more variability in the size and shape of the bipoles which is noticeable in the broadening of the peak distribution of the poloidal and radial energy distribution $\langle B^2\rangle(\ell)$. The differential rotation is a process aligned with the rotation axis, so that the axisymmetry increases with higher differential rotation. 

The last stellar parameter we analysed is the meridional flow. A higher meridional flow more rapidly transports the bipoles and magnetic features to higher latitudes where the shear of the differential rotation is stronger. The bipoles become more tilted and stretched. They form elongated structures that furl around the star and cause a mixed polarity pattern at higher latitudes. They cancel the old polarity of the global dipole and transport the new polarity for the next cycle to the poles more quickly. The poloidal energy $\langle B^2_{\mrm{pol}}\rangle$ decreases for the low $\ell$-modes due to these effects. 
The higher meridional flow pushes the bipoles to such high latitudes that, next to the quadrupolar, also other low to mid even $\ell$-modes carry a high fraction of toroidal and azimuthal field. The even toroidal $\ell$-modes are able to display the polarity switch across the equator, while the odd toroidal modes show the same polarity pattern on both hemispheres. The higher latitudes of the bipoles is best represented by a combination of several even toroidal spherical harmonic modes, where the quadrupolar $\ell = 2$ mode is no longer necessarily the most toroidal one. The bipoles become stretched and tilted due to stronger shearing effects of the differential rotation at the higher latitudes. This causes an increase of the toroidal energy $\langle B^2_{\mrm{tor}}\rangle$ and of $f_{\mrm{azi,tor}}$ for the higher $\ell$-modes. 
The fraction of toroidal axisymmetric field $f_{\mrm{tor,axi}}$ is much lower for the higher meridional flow simulations compared to the solar meridional flow simulations while the fraction of poloidal axisymmetric field $f_{\mrm{pol,axi}}$ remains similar. This effect is especially strong for the dipolar mode. This decreases of the dipolar toroidal axisymmetric field happens as all the magnetic features are dragged to higher latitudes so that nearly no flux is left near the equator to support the dipolar toroidal axisymmetric field. For the solar meridional flow simulations magnetic features emerge and stay much closer to the equator and support the dipolar toroidal axisymmetric field. Moreover, the stretched mixed polarities structures at the higher latitudes add a global non-axisymmertric toroidal structure. 
The meridional energy $\langle B^2_{\mrm{mer}}\rangle$ increases at the cost of the radial energy $\langle B^2_{\mrm{rad}}\rangle$ due to the stronger tilt of the bipoles and their stretch in the meridional direction. The meridional energy distribution $\langle B^2_{\mrm{mer}}\rangle(\ell)$ is also shifted to lower $\ell$-modes as the meridional magnetic features are elongated and the fraction of meridional toroidal field $f_{\mrm{mer,tor}}$ is increased. 
The higher meridional flow causes again a higher variety of the sizes and shapes for the bipoles, which results again in the broadening the poloidal $\langle B^2_{\mrm{pol}}\rangle(\ell)$ and radial energy peak distribution $\langle B^2_{\mrm{rad}}\rangle(\ell)$.

\subsection{Connecting the simulations with the observations}
\label{subsec:ConnectSimandObs}

Some trends found from the simulations for the variation of the large-scale magnetic field topology with the stellar parameters seem to agree with observational results or might be observable. Among others \cite{Vidotto2014} showed that many tracers for the magnetic activity, e.g. the averaged large-scale field strength, increase with decreasing rotation period. Our simulations cover a much smaller parameter range than the observations but the magnetic energy increases with rotation as well. The flux emergence rate has the greatest influence on the magnetic energy and might be connected to the magnetic activity level of the stars. 

For an increasing differential rotation we discovered that the poloidal energy decreases while the toroidal energy remains constant. The simulations show therefore an increase of the toroidal fraction with differential rotation. 
The observations display an increase of the toroidal field fraction with stellar mass \cite[Fig.~3]{Donati2009}. 
Furthermore, \cite{Kitchatinov2011} showed an increase of the differential rotation with increasing stellar mass and effective temperature. The dependency of the differential rotation on the effective temperature is well studied from the observational  \citep{Barnes2005,Reiners2006a,Cameron2007} and from the theoretical side \citep{Kueker2011,Kitchatinov2011}.
Comparing the large-scale magnetic field topology of the mid M-dwarfs \citep{Morin2008a} with the large-scale magnetic field topology of the more massive early M-dwarfs \citep{Donati2008}, we see that the poloidal energy decreases with mass while the toroidal energy remains roughly the same. The fraction of toroidal field increases therefore with mass from the mid to the early M-dwarfs. \cite{Morin2008a} and \cite{Donati2008} showed also that the differential rotation increases by one order of magnitude from the mid to the early M-dwarfs. Our simulations show the same trends for an increasing differential rotation. However, our simulations are based on the Sun and do not represent M-dwarfs.

The meridional flow is not directly observable today. The strongest hint for a higher meridional flow is the decrease of the toroidal axisymmetric fraction while the poloidal axisymmetric fraction remains constant according to our result from the simulations. If one of two solar-analogues shows a weaker toroidal axisymmetric field than the other, could this be a hint of an increased meridional flow. However, this would only be true if both stars show the same solar flux emergence pattern, flux emergence rate and differential rotation.

\section{Summary and Conclusions}
\label{sec:Summary}

We analysed the large- and small-scale magnetic field topology of simulations connecting flux transport models and non-potential coronal evolution models with varying stellar parameters and compared the large-scale magnetic field topology with the observed magnetic field topologies. The main results are the following:

\begin{itemize}

\item The large-scale magnetic field topology of the simulated solar-based stars agrees with the magnetic field topology of the observed solar-like stars.
\item The large-scale field of the simulations is mainly set by the global dipolar field and the  properties of the small-scale flux pattern. The global dipolar field is recovered by the poloidal dominated dipolar mode of the spherical harmonics. The solar small-scale flux emerges at low to mid latitudes with opposite polarities on the two hemispheres and is best recovered on large scales by the toroidal quadrupolar mode. The main spherical harmonic modes of the large-scale field for the simulations are therefore the poloidal dominanted dipolar mode and the quadrupolar mode, which often show the highest fraction of toroidal field. Both are highly axisymmetric. 
\item The magnetic field topology is sensitive to changes in the stellar parameters:

\item An increasing flux emergence rate
  \begin{itemize}
  \item increases the magnetic energy $\langle B^2\rangle$ for all field components.
  \item enhances the toroidal and azimuthal quadrupolar mode relative to the mid and higher $\ell$-modes.
  \end{itemize}

\item An increasing differential rotation
  \begin{itemize}
  \item decreases the poloidal energy $\langle B^2_{\mrm{pol}}\rangle$ and the toroidal energy $\langle B^2_{\mrm{tor}}\rangle$ remains widely constant, which results in an increase of the toroidal field fraction $f_{\mrm{tor}}$.
  \item increases the axisymmetry.
  \end{itemize}
  
\item An increasing meridional flow
  \begin{itemize}
  \item enhances the low even toroidal spherical harmonic modes to mimic the higher latitude position of the magnetic features.
  \item decreases the fraction of the toroidal axisymmetric field $f_{\mrm{tor,axi}}$, especially for the dipolar mode, while the fraction of the poloidal axisymmetric field $f_{\mrm{pol,axi}}$ is unchanged.
  \end{itemize}
  
\item The higher spherical harmonic modes are sensitive to the length scale of the emerging star spots. They show a low fraction of axi\-symmetric field and a widely fixed ratio between the field components that is less affected by the stellar parameters.
\item An increase of the stellar parameters: flux emergence rate, differential rotation and meridional flow cause a greater variation in the size and shape of the magnetic features on the stellar surface. This broadens the poloidal $\langle B^2_{\mrm{pol}}\rangle(\ell)$ and the radial energy distribution $\langle B^2_{\mrm{rad}}\rangle(\ell)$.

\end{itemize}

The results show that simulations connecting photospheric flux transport models with non-potential coronal evolution models are an important tool to understand stellar magnetic field observations. The magnetic field topology, especially the large-scale magnetic field topology, is sensitive to the flux emergence pattern. The simulations we have analysed are based on the observed solar flux emergence pattern between 2000, January and 2001, January. The Sun was in the maximum activity phase of its activity cycle at that time. The dependencies we have discovered of the stellar parameters: flux emergence rate, differential rotation and meridional flow are evaluated for stars with a solar flux emergence pattern of a maximum activity phase. Further examinations regarding the influence of the different flux emergence patterns and cycle phases are recommended to derive more conclusive hints for observable trends. Also a wider parameter range of analysed stellar parameters is necessary to connect securely the trends we have found with the observational trends. The simulations cover only a small range of stellar parameters compared with observations. However, some of the general trends are most likely to be independent of the flux emergence pattern e.g. the increase of magnetic flux with flux emergence  rate and the decrease of the poloidal field and the resultant increase of the toroidal fraction with higher differential rotation.

Additionally, we found hints of similar trends for the solar-based simulations compared to the observations. Both show an increase of the magnetic energy with decreasing rotation period which might be connected to the flux emergence rate. The observations show further an increase of the toroidal field fraction with mass. This is similar to the trends we discover for increasing differential rotation (which is believed to increase with stellar mass).

\section*{Acknowledgements}

LTL acknowledges support from the Scottish Universities Physics Alliance (SUPA) prize studentship and the University of St Andrews Higgs studentship. MMJ acknowledges the support of the Science \& Technology Facilities Council (STFC) (ST/M001296/1).




\bibliographystyle{mnras}
\bibliography{Lehmann2017Connecting} 

\begin{thebibliography}{}
\makeatletter
\relax
\def\mn@urlcharsother{\let\do\@makeother \do\$\do\&\do\#\do\^\do\_\do\%\do\~}
\def\mn@doi{\begingroup\mn@urlcharsother \@ifnextchar [ {\mn@doi@}
  {\mn@doi@[]}}
\def\mn@doi@[#1]#2{\def\@tempa{#1}\ifx\@tempa\@empty \href
  {http://dx.doi.org/#2} {doi:#2}\else \href {http://dx.doi.org/#2} {#1}\fi
  \endgroup}
\def\mn@eprint#1#2{\mn@eprint@#1:#2::\@nil}
\def\mn@eprint@arXiv#1{\href {http://arxiv.org/abs/#1} {{\tt arXiv:#1}}}
\def\mn@eprint@dblp#1{\href {http://dblp.uni-trier.de/rec/bibtex/#1.xml}
  {dblp:#1}}
\def\mn@eprint@#1:#2:#3:#4\@nil{\def\@tempa {#1}\def\@tempb {#2}\def\@tempc
  {#3}\ifx \@tempc \@empty \let \@tempc \@tempb \let \@tempb \@tempa \fi \ifx
  \@tempb \@empty \def\@tempb {arXiv}\fi \@ifundefined
  {mn@eprint@\@tempb}{\@tempb:\@tempc}{\expandafter \expandafter \csname
  mn@eprint@\@tempb\endcsname \expandafter{\@tempc}}}

\bibitem[\protect\citeauthoryear{Arzoumanian, Jardine, Donati, Morin  \&
  Johnstone}{Arzoumanian et~al.}{2011}]{Arzoumanian2011}
Arzoumanian D.,  Jardine M.,  Donati J.-F.,  Morin J.,   Johnstone C.,  2011,
  \mn@doi [\mnras] {10.1111/j.1365-2966.2010.17623.x}, 410, 2472

\bibitem[\protect\citeauthoryear{Babcock}{Babcock}{1961}]{Babcock1961}
Babcock H.~W.,  1961, \mn@doi [\apj] {10.1086/147060}, 133, 572

\bibitem[\protect\citeauthoryear{Baklanova \& Plachinda}{Baklanova \&
  Plachinda}{2015}]{Baklanova2015}
Baklanova D.,  Plachinda S.,  2015, \mn@doi [Adv. Space Res.]
  {10.1016/j.asr.2014.05.029}, 55, 817

\bibitem[\protect\citeauthoryear{{Barnes}, {Collier Cameron}, {Donati},
  {James}, {Marsden}  \& {Petit}}{{Barnes} et~al.}{2005}]{Barnes2005}
{Barnes} J.~R.,  {Collier Cameron} A.,  {Donati} J.-F.,  {James} D.~J.,
  {Marsden} S.~C.,   {Petit} P.,  2005, \mn@doi [\mnras]
  {10.1111/j.1745-3933.2005.08587.x}, \href
  {http://adsabs.harvard.edu/abs/2005MNRAS.357L...1B} {357, L1}

\bibitem[\protect\citeauthoryear{Baumann, Schmitt, Sch{\"u}ssler  \&
  Solanki}{Baumann et~al.}{2004}]{Baumann2004}
Baumann I.,  Schmitt D.,  Sch{\"u}ssler M.,   Solanki S.~K.,  2004, \mn@doi
  [\aap] {10.1051/0004-6361:20048024}, 426, 1075

\bibitem[\protect\citeauthoryear{{Bonanno}}{{Bonanno}}{2016}]{Bonanno2016}
{Bonanno} A.,  2016, \mn@doi [\apjl] {10.3847/2041-8213/833/2/L22}, \href
  {http://adsabs.harvard.edu/abs/2016ApJ...833L..22B} {833, L22}

\bibitem[\protect\citeauthoryear{Boro~Saikia, Jeffers, Petit, Marsden, Morin
  \& Folsom}{Boro~Saikia et~al.}{2015}]{BoroSaikia2015}
Boro~Saikia S.,  Jeffers S.~V.,  Petit P.,  Marsden S.,  Morin J.,   Folsom
  C.~P.,  2015, \mn@doi [\aap] {10.1051/0004-6361/201424096}, 573, A17

\bibitem[\protect\citeauthoryear{{Cameron} \& {Sch{\"u}ssler}}{{Cameron} \&
  {Sch{\"u}ssler}}{2015}]{Cameron2015}
{Cameron} R.,  {Sch{\"u}ssler} M.,  2015, \mn@doi [Science]
  {10.1126/science.1261470}, \href
  {http://adsabs.harvard.edu/abs/2015Sci...347.1333C} {347, 1333}

\bibitem[\protect\citeauthoryear{Chandrasekhar}{Chandrasekhar}{1961}]{Chandrasekhar1961}
Chandrasekhar S.,  1961, Hydrodynamic and Hydromagnetic Stability

\bibitem[\protect\citeauthoryear{{Collier Cameron}}{{Collier
  Cameron}}{2007}]{Cameron2007}
{Collier Cameron} A.,  2007, \mn@doi [Astronomische Nachrichten]
  {10.1002/asna.200710880}, \href
  {http://adsabs.harvard.edu/abs/2007AN....328.1030C} {328, 1030}

\bibitem[\protect\citeauthoryear{DeRosa, Brun  \& Hoeksema}{DeRosa
  et~al.}{2012}]{DeRosa2012}
DeRosa M.~L.,  Brun A.~S.,   Hoeksema J.~T.,  2012, \mn@doi [\apj]
  {10.1088/0004-637X/757/1/96}, 757, 96

\bibitem[\protect\citeauthoryear{{DeVore}, {Sheeley}, {Boris}, {Young}  \&
  {Harvey}}{{DeVore} et~al.}{1985}]{DeVore1985}
{DeVore} C.~R.,  {Sheeley} Jr. N.~R.,  {Boris} J.~P.,  {Young} Jr. T.~R.,
  {Harvey} K.~L.,  1985, \mn@doi [\solphys] {10.1007/BF00154036}, \href
  {http://adsabs.harvard.edu/abs/1985SoPh..102...41D} {102, 41}

\bibitem[\protect\citeauthoryear{{Donahue}, {Saar}  \& {Baliunas}}{{Donahue}
  et~al.}{1996}]{Donahue1996}
{Donahue} R.~A.,  {Saar} S.~H.,   {Baliunas} S.~L.,  1996, \mn@doi [\apj]
  {10.1086/177517}, \href {http://adsabs.harvard.edu/abs/1996ApJ...466..384D}
  {466, 384}

\bibitem[\protect\citeauthoryear{Donati \& Brown}{Donati \&
  Brown}{1997}]{Donati1997}
Donati J.-F.,  Brown S.~F.,  1997, \aap, 326, 1135

\bibitem[\protect\citeauthoryear{Donati \& Landstreet}{Donati \&
  Landstreet}{2009}]{Donati2009}
Donati J.-F.,  Landstreet J.,  2009, \mn@doi [Annu. Rev. Astron. Astrophys.]
  {10.1146/annurev-astro-082708-101833}, 47, 333

\bibitem[\protect\citeauthoryear{Donati et~al.,}{Donati
  et~al.}{2003}]{Donati2003}
Donati J.-F.,  et~al., 2003, \mn@doi [\mnras]
  {10.1046/j.1365-2966.2003.07031.x}, 345, 1145

\bibitem[\protect\citeauthoryear{Donati, Forveille, Collier~Cameron, Barnes,
  Delfosse, Jardine  \& Valenti}{Donati et~al.}{2006a}]{Donati2006}
Donati J.-F.,  Forveille T.,  Collier~Cameron A.,  Barnes J.~R.,  Delfosse X.,
  Jardine M.~M.,   Valenti J.~A.,  2006a, \mn@doi [Science]
  {10.1126/science.1121102}, 311, 633

\bibitem[\protect\citeauthoryear{Donati et~al.,}{Donati
  et~al.}{2006b}]{Donati2006a}
Donati J.-F.,  et~al., 2006b, \mn@doi [\mnras]
  {10.1111/j.1365-2966.2006.10558.x}, 370, 629

\bibitem[\protect\citeauthoryear{Donati et~al.,}{Donati
  et~al.}{2008}]{Donati2008}
Donati J.-F.,  et~al., 2008, \mn@doi [\mnras]
  {10.1111/j.1365-2966.2008.13799.x}, 390, 545

\bibitem[\protect\citeauthoryear{Elsasser}{Elsasser}{1946}]{Elsasser1946}
Elsasser W.~M.,  1946, \mn@doi [Physical Review] {10.1103/PhysRev.69.106}, 69,
  106

\bibitem[\protect\citeauthoryear{{Fan}}{{Fan}}{2001}]{Fan2001}
{Fan} Y.,  2001, \mn@doi [\apjl] {10.1086/320935}, \href
  {http://adsabs.harvard.edu/abs/2001ApJ...554L.111F} {554, L111}

\bibitem[\protect\citeauthoryear{Fares et~al.,}{Fares et~al.}{2009}]{Fares2009}
Fares R.,  et~al., 2009, \mn@doi [\mnras] {10.1111/j.1365-2966.2009.15303.x},
  398, 1383

\bibitem[\protect\citeauthoryear{Fares et~al.,}{Fares et~al.}{2010}]{Fares2010}
Fares R.,  et~al., 2010, \mn@doi [\mnras] {10.1111/j.1365-2966.2010.16715.x},
  406, 409

\bibitem[\protect\citeauthoryear{Fares et~al.,}{Fares et~al.}{2012}]{Fares2012}
Fares R.,  et~al., 2012, \mn@doi [\mnras] {10.1111/j.1365-2966.2012.20780.x},
  423, 1006

\bibitem[\protect\citeauthoryear{Fares, Moutou, Donati, Catala, Shkolnik,
  Jardine, Cameron  \& Deleuil}{Fares et~al.}{2013}]{Fares2013}
Fares R.,  Moutou C.,  Donati J.-F.,  Catala C.,  Shkolnik E.~L.,  Jardine
  M.~M.,  Cameron A.~C.,   Deleuil M.,  2013, \mn@doi [\mnras]
  {10.1093/mnras/stt1386}, 435, 1451

\bibitem[\protect\citeauthoryear{Folsom et~al.,}{Folsom
  et~al.}{2016}]{Folsom2016}
Folsom C.~P.,  et~al., 2016, \mn@doi [\mnras] {10.1093/mnras/stv2924}, 457, 580

\bibitem[\protect\citeauthoryear{Gibb, Mackay, Jardine  \& Yeates}{Gibb
  et~al.}{2016}]{Gibb2016}
Gibb G. P.~S.,  Mackay D.~H.,  Jardine M.~M.,   Yeates A.~R.,  2016, \mn@doi
  [\mnras] {10.1093/mnras/stv2920}, 456, 3624

\bibitem[\protect\citeauthoryear{G{\"u}del}{G{\"u}del}{2007}]{Guedel2007}
G{\"u}del M.,  2007, \mn@doi [Living Reviews in Solar Physics]
  {10.12942/lrsp-2007-3}, 4

\bibitem[\protect\citeauthoryear{{Hale}, {Ellerman}, {Nicholson}  \&
  {Joy}}{{Hale} et~al.}{1919}]{Hale1919}
{Hale} G.~E.,  {Ellerman} F.,  {Nicholson} S.~B.,   {Joy} A.~H.,  1919, \mn@doi
  [\apj] {10.1086/142452}, \href
  {http://adsabs.harvard.edu/abs/1919ApJ....49..153H} {49, 153}

\bibitem[\protect\citeauthoryear{Hartmann \& Noyes}{Hartmann \&
  Noyes}{1987}]{Hartmann1987}
Hartmann L.~W.,  Noyes R.~W.,  1987, \mn@doi [Annual Review of Astronomy and
  Astrophysics] {10.1146/annurev.aa.25.090187.001415}, 25, 271

\bibitem[\protect\citeauthoryear{H{\'e}brard, Donati, Delfosse, Morin, Moutou
  \& Boisse}{H{\'e}brard et~al.}{2016}]{Hebrard2016}
H{\'e}brard {\'E}.~M.,  Donati J.-F.,  Delfosse X.,  Morin J.,  Moutou C.,
  Boisse I.,  2016, \mn@doi [\mnras] {10.1093/mnras/stw1346}, 461, 1465

\bibitem[\protect\citeauthoryear{{Hill}, {Carmona}, {Donati}, {Hussain},
  {Gregory}, {Alencar}, {Bouvier}  \& {the MaTYSSE collaboration}}{{Hill}
  et~al.}{2017}]{Hill2017}
{Hill} C.~A.,  {Carmona} A.,  {Donati} J.-F.,  {Hussain} G.~A.~J.,  {Gregory}
  S.~G.,  {Alencar} S.~H.~P.,  {Bouvier} J.,   {the MaTYSSE collaboration}
  2017, preprint, \href {http://adsabs.harvard.edu/abs/2017arXiv170809693H} {}
  (\mn@eprint {arXiv} {1708.09693})

\bibitem[\protect\citeauthoryear{{Holzwarth}, {Mackay}  \&
  {Jardine}}{{Holzwarth} et~al.}{2007}]{Holzwarth2007}
{Holzwarth} V.,  {Mackay} D.~H.,   {Jardine} M.,  2007, \mn@doi [Astronomische
  Nachrichten] {10.1002/asna.200710854}, \href
  {http://adsabs.harvard.edu/abs/2007AN....328.1108H} {328, 1108}

\bibitem[\protect\citeauthoryear{{Hung}, {Brun}, {Fournier}, {Jouve},
  {Talagrand}  \& {Zakari}}{{Hung} et~al.}{2017}]{Hung2017}
{Hung} C.~P.,  {Brun} A.~S.,  {Fournier} A.,  {Jouve} L.,  {Talagrand} O.,
  {Zakari} M.,  2017, \mn@doi [\apj] {10.3847/1538-4357/aa91d1}, \href
  {http://adsabs.harvard.edu/abs/2017ApJ...849..160H} {849, 160}

\bibitem[\protect\citeauthoryear{{I{\c s}{\i}k}, {Schmitt}  \&
  {Sch{\"u}ssler}}{{I{\c s}{\i}k} et~al.}{2011}]{Isik2011}
{I{\c s}{\i}k} E.,  {Schmitt} D.,   {Sch{\"u}ssler} M.,  2011, \mn@doi [\aap]
  {10.1051/0004-6361/201014501}, \href
  {http://adsabs.harvard.edu/abs/2011A%26A...528A.135I} {528, A135}

\bibitem[\protect\citeauthoryear{Jeffers, Petit, Marsden, Morin, Donati  \&
  Folsom}{Jeffers et~al.}{2014}]{Jeffers2014}
Jeffers S.~V.,  Petit P.,  Marsden S.~C.,  Morin J.,  Donati J.-F.,   Folsom
  C.~P.,  2014, \mn@doi [\aap] {10.1051/0004-6361/201423725}, 569, A79

\bibitem[\protect\citeauthoryear{Jiang, Cameron, Schmitt  \& I{\c
  s}{\i}k}{Jiang et~al.}{2013}]{Jiang2013}
Jiang J.,  Cameron R.~H.,  Schmitt D.,   I{\c s}{\i}k E.,  2013, \mn@doi [\aap]
  {10.1051/0004-6361/201321145}, 553, A128

\bibitem[\protect\citeauthoryear{Johnstone, Jardine  \& Mackay}{Johnstone
  et~al.}{2010}]{Johnstone2010}
Johnstone C.,  Jardine M.,   Mackay D.~H.,  2010, \mn@doi [\mnras]
  {10.1111/j.1365-2966.2010.16298.x}, 404, 101

\bibitem[\protect\citeauthoryear{Johnstone, Jardine, Gregory, Donati  \&
  Hussain}{Johnstone et~al.}{2014}]{Johnstone2014}
Johnstone C.~P.,  Jardine M.,  Gregory S.~G.,  Donati J.-F.,   Hussain G.,
  2014, \mn@doi [\mnras] {10.1093/mnras/stt2107}, 437, 3202

\bibitem[\protect\citeauthoryear{{Karak} \& {Miesch}}{{Karak} \&
  {Miesch}}{2017}]{Karak2017}
{Karak} B.~B.,  {Miesch} M.,  2017, \mn@doi [\apj] {10.3847/1538-4357/aa8636},
  \href {http://adsabs.harvard.edu/abs/2017ApJ...847...69K} {847, 69}

\bibitem[\protect\citeauthoryear{{Kitchatinov} \& {Olemskoy}}{{Kitchatinov} \&
  {Olemskoy}}{2011}]{Kitchatinov2011}
{Kitchatinov} L.~L.,  {Olemskoy} S.~V.,  2011, \mn@doi [\mnras]
  {10.1111/j.1365-2966.2010.17737.x}, \href
  {http://adsabs.harvard.edu/abs/2011MNRAS.411.1059K} {411, 1059}

\bibitem[\protect\citeauthoryear{{K{\"u}ker} \& {R{\"u}diger}}{{K{\"u}ker} \&
  {R{\"u}diger}}{2011}]{Kueker2011}
{K{\"u}ker} M.,  {R{\"u}diger} G.,  2011, \mn@doi [Astronomische Nachrichten]
  {10.1002/asna.201111628}, \href
  {http://adsabs.harvard.edu/abs/2011AN....332..933K} {332, 933}

\bibitem[\protect\citeauthoryear{Lang, Jardine, Morin, Donati, Jeffers, Vidotto
   \& Fares}{Lang et~al.}{2014}]{Lang2014}
Lang P.,  Jardine M.,  Morin J.,  Donati J.-F.,  Jeffers S.,  Vidotto A.~A.,
  Fares R.,  2014, \mn@doi [\mnras] {10.1093/mnras/stu091}, 439, 2122

\bibitem[\protect\citeauthoryear{Larmor}{Larmor}{1919}]{Larmor1919}
Larmor J.,  1919, Rep Brit Assoc Adv Sci, 159, 412

\bibitem[\protect\citeauthoryear{Lehmann, K{\"u}nstler, Carroll  \&
  Strassmeier}{Lehmann et~al.}{2015}]{Lehmann2015}
Lehmann L.~T.,  K{\"u}nstler A.,  Carroll T.~A.,   Strassmeier K.~G.,  2015,
  \mn@doi [Astron. Nachrichten] {10.1002/asna.201412162}, 336, 258

\bibitem[\protect\citeauthoryear{Lehmann et~al.,}{Lehmann
  et~al.}{2017}]{Lehmann2017}
Lehmann L.~T.,  et~al., 2017, \mn@doi [\mnras] {10.1093/mnrasl/slw225}, 466,
  L24

\bibitem[\protect\citeauthoryear{{Leighton}}{{Leighton}}{1964}]{Leighton1964}
{Leighton} R.~B.,  1964, \mn@doi [\apj] {10.1086/148058}, \href
  {http://adsabs.harvard.edu/abs/1964ApJ...140.1547L} {140, 1547}

\bibitem[\protect\citeauthoryear{Leighton}{Leighton}{1969}]{Leighton1969}
Leighton R.~B.,  1969, \mn@doi [\apj] {10.1086/149943}, 156, 1

\bibitem[\protect\citeauthoryear{Mackay \& Yeates}{Mackay \&
  Yeates}{2012}]{Mackay2012}
Mackay D.,  Yeates A.,  2012, \mn@doi [Living Reviews in Solar Physics]
  {10.12942/lrsp-2012-6}, 9

\bibitem[\protect\citeauthoryear{Mackay \& {van Ballegooijen}}{Mackay \& {van
  Ballegooijen}}{2006}]{Mackay2006}
Mackay D.~H.,  {van Ballegooijen} A.~A.,  2006, \mn@doi [\apj]
  {10.1086/500425}, 641, 577

\bibitem[\protect\citeauthoryear{Mackay, Jardine, Collier~Cameron, Donati  \&
  Hussain}{Mackay et~al.}{2004}]{Mackay2004}
Mackay D.~H.,  Jardine M.,  Collier~Cameron A.,  Donati J.-F.,   Hussain G.
  A.~J.,  2004, \mn@doi [\mnras] {10.1111/j.1365-2966.2004.08233.x}, 354, 737

\bibitem[\protect\citeauthoryear{Marsden, Donati, Semel, Petit  \&
  Carter}{Marsden et~al.}{2006}]{Marsden2006}
Marsden S.~C.,  Donati J.-F.,  Semel M.,  Petit P.,   Carter B.~D.,  2006,
  \mn@doi [\mnras] {10.1111/j.1365-2966.2006.10503.x}, 370, 468

\bibitem[\protect\citeauthoryear{{Marsden} et~al.,}{{Marsden}
  et~al.}{2011}]{Marsden2011}
{Marsden} S.~C.,  et~al., 2011, \mn@doi [\mnras]
  {10.1111/j.1365-2966.2011.18272.x}, \href
  {http://adsabs.harvard.edu/abs/2011MNRAS.413.1939M} {413, 1939}

\bibitem[\protect\citeauthoryear{Marsden et~al.,}{Marsden
  et~al.}{2014}]{Marsden2014}
Marsden S.~C.,  et~al., 2014, \mn@doi [\mnras] {10.1093/mnras/stu1663}, 444,
  3517

\bibitem[\protect\citeauthoryear{{Mekkaden}}{{Mekkaden}}{1985}]{Mekkaden1985}
{Mekkaden} M.~V.,  1985, \mn@doi [\apss] {10.1007/BF00650163}, \href
  {http://adsabs.harvard.edu/abs/1985Ap%26SS.117..381M} {117, 381}

\bibitem[\protect\citeauthoryear{{Middelkoop}}{{Middelkoop}}{1981}]{Middelkoop1981}
{Middelkoop} F.,  1981, \aap, \href
  {http://adsabs.harvard.edu/abs/1981A%26A...101..295M} {101, 295}

\bibitem[\protect\citeauthoryear{{Miesch} \& {Teweldebirhan}}{{Miesch} \&
  {Teweldebirhan}}{2016}]{Miesch2016}
{Miesch} M.~S.,  {Teweldebirhan} K.,  2016, \mn@doi [Advances in Space
  Research] {10.1016/j.asr.2016.02.018}, \href
  {http://adsabs.harvard.edu/abs/2016AdSpR..58.1571M} {58, 1571}

\bibitem[\protect\citeauthoryear{Morin et~al.,}{Morin
  et~al.}{2008a}]{Morin2008}
Morin J.,  et~al., 2008a, \mn@doi [\mnras] {10.1111/j.1365-2966.2007.12709.x},
  384, 77

\bibitem[\protect\citeauthoryear{Morin et~al.,}{Morin
  et~al.}{2008b}]{Morin2008a}
Morin J.,  et~al., 2008b, \mn@doi [\mnras] {10.1111/j.1365-2966.2008.13809.x},
  390, 567

\bibitem[\protect\citeauthoryear{Morin, Donati, Petit, Delfosse, Forveille  \&
  Jardine}{Morin et~al.}{2010}]{Morin2010}
Morin J.,  Donati J.-F.,  Petit P.,  Delfosse X.,  Forveille T.,   Jardine
  M.~M.,  2010, \mn@doi [\mnras] {10.1111/j.1365-2966.2010.17101.x}, 407, 2269

\bibitem[\protect\citeauthoryear{{Morris}, {Hebb}, {Davenport}, {Rohn}  \&
  {Hawley}}{{Morris} et~al.}{2017}]{Morris2017}
{Morris} B.~M.,  {Hebb} L.,  {Davenport} J.~R.~A.,  {Rohn} G.,   {Hawley}
  S.~L.,  2017, \mn@doi [\apj] {10.3847/1538-4357/aa8555}, \href
  {http://adsabs.harvard.edu/abs/2017ApJ...846...99M} {846, 99}

\bibitem[\protect\citeauthoryear{{Pallavicini}, {Golub}, {Rosner}, {Vaiana},
  {Ayres}  \& {Linsky}}{{Pallavicini} et~al.}{1981}]{Pallavicini1981}
{Pallavicini} R.,  {Golub} L.,  {Rosner} R.,  {Vaiana} G.~S.,  {Ayres} T.,
  {Linsky} J.~L.,  1981, \mn@doi [\apj] {10.1086/159152}, \href
  {http://adsabs.harvard.edu/abs/1981ApJ...248..279P} {248, 279}

\bibitem[\protect\citeauthoryear{Parker}{Parker}{1955}]{Parker1955}
Parker E.~N.,  1955, \mn@doi [\apj] {10.1086/146087}, 122, 293

\bibitem[\protect\citeauthoryear{{Petit}, {Donati}  \& {Collier
  Cameron}}{{Petit} et~al.}{2002}]{Petit2002}
{Petit} P.,  {Donati} J.-F.,   {Collier Cameron} A.,  2002, \mn@doi [\mnras]
  {10.1046/j.1365-8711.2002.05529.x}, \href
  {http://adsabs.harvard.edu/abs/2002MNRAS.334..374P} {334, 374}

\bibitem[\protect\citeauthoryear{Petit et~al.,}{Petit et~al.}{2008}]{Petit2008}
Petit P.,  et~al., 2008, \mn@doi [\mnras] {10.1111/j.1365-2966.2008.13411.x},
  388, 80

\bibitem[\protect\citeauthoryear{{Priest}}{{Priest}}{1982}]{Priest1982}
{Priest} E.~R.,  1982, {Solar magneto-hydrodynamics}

\bibitem[\protect\citeauthoryear{{Reiners}}{{Reiners}}{2006}]{Reiners2006a}
{Reiners} A.,  2006, \mn@doi [\aap] {10.1051/0004-6361:20053911}, \href
  {http://adsabs.harvard.edu/abs/2006A%26A...446..267R} {446, 267}

\bibitem[\protect\citeauthoryear{Reiners}{Reiners}{2012}]{Reiners2012}
Reiners A.,  2012, \mn@doi [Living Reviews in Solar Physics]
  {10.12942/lrsp-2012-1}, 9, 1

\bibitem[\protect\citeauthoryear{Reiners \& Basri}{Reiners \&
  Basri}{2006}]{Reiners2006}
Reiners A.,  Basri G.,  2006, \mn@doi [\apj] {10.1086/503324}, 644, 497

\bibitem[\protect\citeauthoryear{Robinson, Worden  \& Harvey}{Robinson
  et~al.}{1980}]{Robinson1980}
Robinson R.~D.,  Worden S.~P.,   Harvey J.~W.,  1980, \mn@doi [\apjl]
  {10.1086/183217}, 236, L155

\bibitem[\protect\citeauthoryear{Ros{\'e}n, Kochukhov, Hackman  \&
  Lehtinen}{Ros{\'e}n et~al.}{2016}]{Rosen2016}
Ros{\'e}n L.,  Kochukhov O.,  Hackman T.,   Lehtinen J.,  2016,
  arXiv:1605.03026 [astro-ph]

\bibitem[\protect\citeauthoryear{Saar}{Saar}{1988}]{Saar1988}
Saar S.~H.,  1988, \mn@doi [\apj] {10.1086/165907}, 324, 441

\bibitem[\protect\citeauthoryear{Saar}{Saar}{1996}]{Saar1996}
Saar S.~H.,  1996. p.~237

\bibitem[\protect\citeauthoryear{{Scalia}, {Leone}, {Gangi}, {Giarrusso}  \&
  {Stift}}{{Scalia} et~al.}{2017}]{Scalia2017}
{Scalia} C.,  {Leone} F.,  {Gangi} M.,  {Giarrusso} M.,   {Stift} M.~J.,  2017,
  preprint, \href {http://adsabs.harvard.edu/abs/2017arXiv170805196S} {}
  (\mn@eprint {arXiv} {1708.05196})

\bibitem[\protect\citeauthoryear{{Schrijver} \& {Title}}{{Schrijver} \&
  {Title}}{2001}]{Schrijver2001}
{Schrijver} C.~J.,  {Title} A.~M.,  2001, \mn@doi [\apj] {10.1086/320237},
  \href {http://adsabs.harvard.edu/abs/2001ApJ...551.1099S} {551, 1099}

\bibitem[\protect\citeauthoryear{See et~al.,}{See et~al.}{2015}]{See2015}
See V.,  et~al., 2015, \mn@doi [\mnras] {10.1093/mnras/stv1925}, 453, 4301

\bibitem[\protect\citeauthoryear{Semel}{Semel}{1989}]{Semel1989}
Semel M.,  1989, \aap, 225, 456

\bibitem[\protect\citeauthoryear{Skumanich}{Skumanich}{1972}]{Skumanich1972}
Skumanich A.,  1972, \mn@doi [\apj] {10.1086/151310}, 171, 565

\bibitem[\protect\citeauthoryear{Vidotto}{Vidotto}{2016}]{Vidotto2016a}
Vidotto A.~A.,  2016, \mn@doi [\mnras] {10.1093/mnras/stw758}, 459, 1533

\bibitem[\protect\citeauthoryear{Vidotto et~al.,}{Vidotto
  et~al.}{2014}]{Vidotto2014}
Vidotto A.~A.,  et~al., 2014, \mn@doi [\mnras] {10.1093/mnras/stu728}, 441,
  2361

\bibitem[\protect\citeauthoryear{Vidotto et~al.,}{Vidotto
  et~al.}{2016}]{Vidotto2016}
Vidotto A.~A.,  et~al., 2016, \mn@doi [\mnras] {10.1093/mnrasl/slv147}, 455,
  L52

\bibitem[\protect\citeauthoryear{Waite, Marsden, Carter, Hart, Donati,
  Ram{\'\i}rez~V{\'e}lez, Semel  \& Dunstone}{Waite et~al.}{2011}]{Waite2011}
Waite I.~A.,  Marsden S.~C.,  Carter B.~D.,  Hart R.,  Donati J.-F.,
  Ram{\'\i}rez~V{\'e}lez J.~C.,  Semel M.,   Dunstone N.,  2011, \mn@doi
  [\mnras] {10.1111/j.1365-2966.2011.18366.x}, 413, 1949

\bibitem[\protect\citeauthoryear{{Walter} \& {Bowyer}}{{Walter} \&
  {Bowyer}}{1981}]{Walter1981}
{Walter} F.~M.,  {Bowyer} S.,  1981, \mn@doi [\apj] {10.1086/158842}, \href
  {http://adsabs.harvard.edu/abs/1981ApJ...245..671W} {245, 671}

\bibitem[\protect\citeauthoryear{Wang, Nash  \& Sheeley}{Wang
  et~al.}{1989}]{Wang1989}
Wang Y.-M.,  Nash A.~G.,   Sheeley Jr. N.~R.,  1989, \mn@doi [Science]
  {10.1126/science.245.4919.712}, 245, 712

\bibitem[\protect\citeauthoryear{{Weber} \& {Browning}}{{Weber} \&
  {Browning}}{2016}]{Weber2016}
{Weber} M.~A.,  {Browning} M.~K.,  2016, \mn@doi [\apj]
  {10.3847/0004-637X/827/2/95}, \href
  {http://adsabs.harvard.edu/abs/2016ApJ...827...95W} {827, 95}

\bibitem[\protect\citeauthoryear{Yadav, Christensen, Morin, Gastine, Reiners,
  Poppenhaeger  \& Wolk}{Yadav et~al.}{2015}]{Yadav2015}
Yadav R.~K.,  Christensen U.~R.,  Morin J.,  Gastine T.,  Reiners A.,
  Poppenhaeger K.,   Wolk S.~J.,  2015, \mn@doi [\apjl]
  {10.1088/2041-8205/813/2/L31}, 813, L31

\bibitem[\protect\citeauthoryear{Yeates \& Mackay}{Yeates \&
  Mackay}{2012}]{Yeates2012}
Yeates A.~R.,  Mackay D.~H.,  2012, \mn@doi [\apjl]
  {10.1088/2041-8205/753/2/L34}, 753, L34

\bibitem[\protect\citeauthoryear{{do Nascimento}, {et al., 2014 in}, {Jardine}
  \& {Spruit}}{{do Nascimento} et~al.}{2014}]{doNascimento2014}
{do Nascimento} J.~D.,  {et al., 2014 in} {Petit} P.,  {Jardine} M.,   {Spruit}
  H.~C.,  eds, 2014, {Magnetic Fields throughout Stellar Evolution (IAU S302)}
  IAU Symposium Vol. 302

\bibitem[\protect\citeauthoryear{{van Ballegooijen}, {Priest}  \&
  {Mackay}}{{van Ballegooijen} et~al.}{2000}]{VanBallegooijen2000}
{van Ballegooijen} A.~A.,  {Priest} E.~R.,   {Mackay} D.~H.,  2000, \mn@doi
  [\apj] {10.1086/309265}, \href
  {http://adsabs.harvard.edu/abs/2000ApJ...539..983V} {539, 983}

\makeatother
\end{thebibliography}



\appendix

\section{additional figures}

The following figures provide additional information and results to the previous shown figures.

\begin{figure}
	\includegraphics[width=\columnwidth]{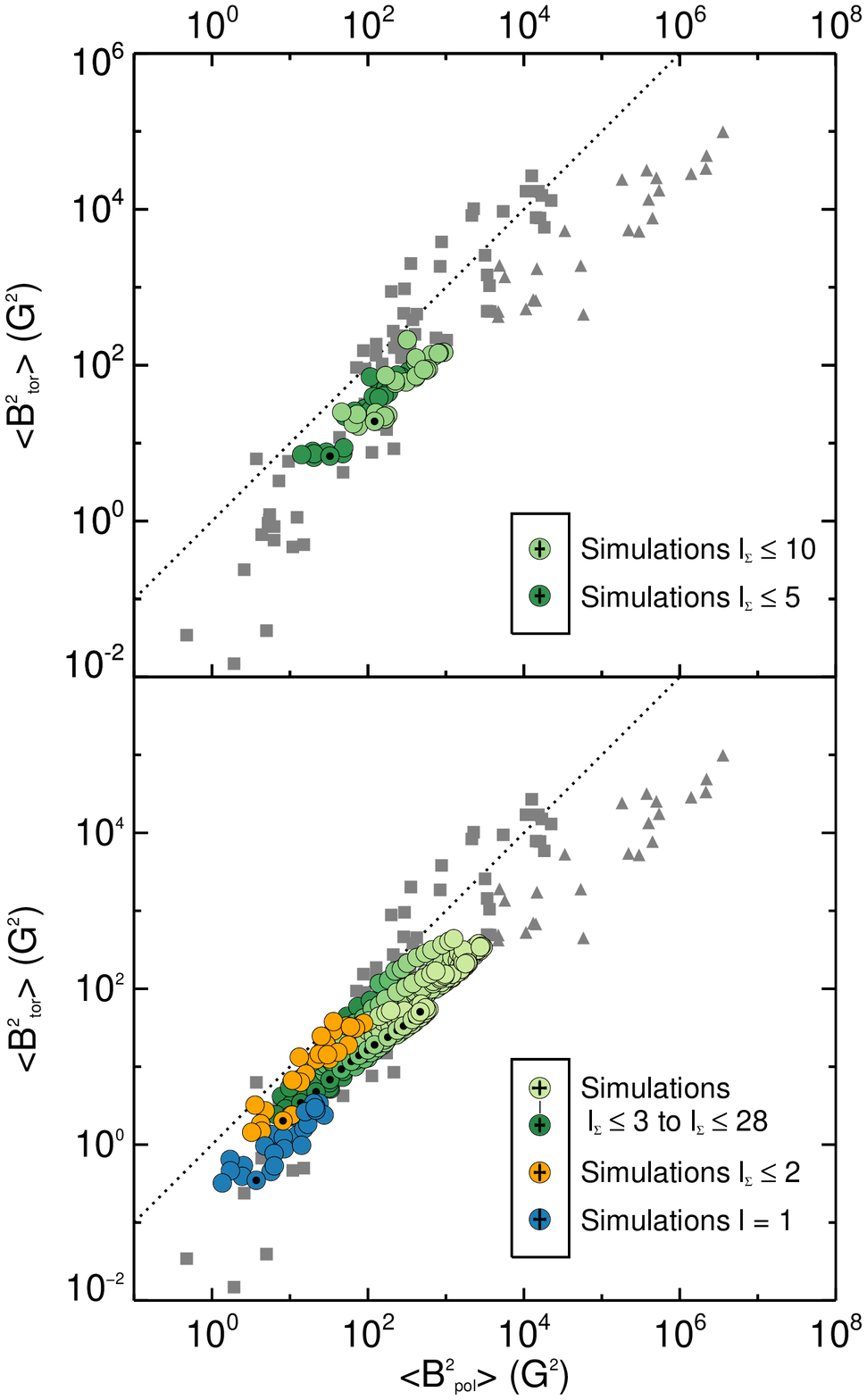}
    \caption{The magnetic energy stored in the poloidal $\langle B^2_{\mathrm{pol}}\rangle$ and the toroidal component $\langle B^2_{\mathrm{tor}}\rangle$ for all simulations including the lower DR, higher DR and higher MF simulations. The higher $\ellsum\ $-modes follow the powerlaw $\langle B^2_{\mathrm{tor}}\rangle \propto \langle B^2_{\mathrm{pol}}\rangle^{0.77\pm0.01}$. The same format as in Fig.~\ref{fig:BtorBpolRepeatori} is used.}
    \label{fig:BtorBpol_All}
\end{figure}

Figure~\ref{fig:BtorBpol_All} shows all analysed simulations together (lower DR, higher DR and higher MF simulations) in the exact same format as in Fig.~\ref{fig:BtorBpolRepeatori}, where only the higher DR simulations are presented. The powerlaw for the higher $\ellsum\ $-modes (greenish symbols) for all simulations together is $\langle B^2_{\mathrm{tor}}\rangle \propto \langle B^2_{\mathrm{pol}}\rangle^{0.77\pm0.01}$.

\begin{figure*}

	\includegraphics[width=2\columnwidth]{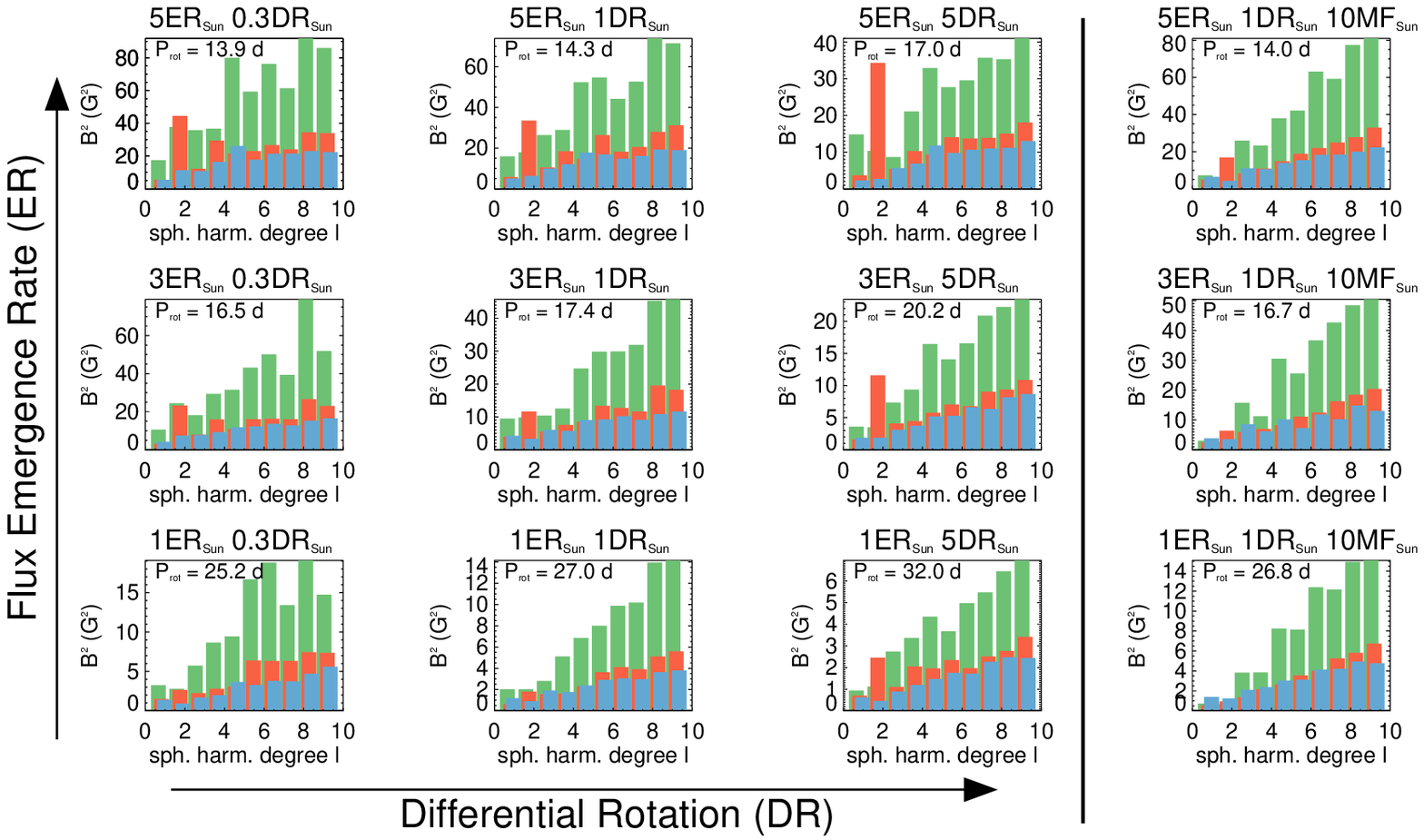}
    \caption{The first ten $\ell$-modes of the energy distribution for the radial $\langle B^2_{\mathrm{rad}}\rangle(\ell)$ (green bars), azimuthal $\langle B^2_{\mathrm{azi}}\rangle(\ell)$ (red bars) and meridional component $\langle B^2_{\mathrm{mer}}\rangle(\ell)$ (blue bars). The y-axis varies for all simulations and the rotation period of the simulated star is displayed in the \textit{top left} corner of each barplot. The same format as in Fig.~\ref{fig:BP_E_RadAziMer} is used.}
    \label{fig:BP_E_RadAziMer_zoom}
\end{figure*}

Figure~\ref{fig:BP_E_RadAziMer_zoom} provides a closer look to the first ten $\ell$-modes of Fig.~\ref{fig:BP_E_RadAziMer}, where all $\ell$-modes $\ell = 1-28$ are shown. Be aware that the y-axis is not longer fixed and varies for all simulations. The rotation period of the simulated star is mentioned in the \textit{top left} corner of each barplot. The strong azimuthal $\ell = 2$ mode is more obvious in this format. Figure~\ref{fig:BP_E_RadAziMer_zoom} allows further an easier comparison of the magnetic energy distribution for the large-scale field with results from observations published in the literature.

\begin{figure*}
	\includegraphics[width=1.33\columnwidth]{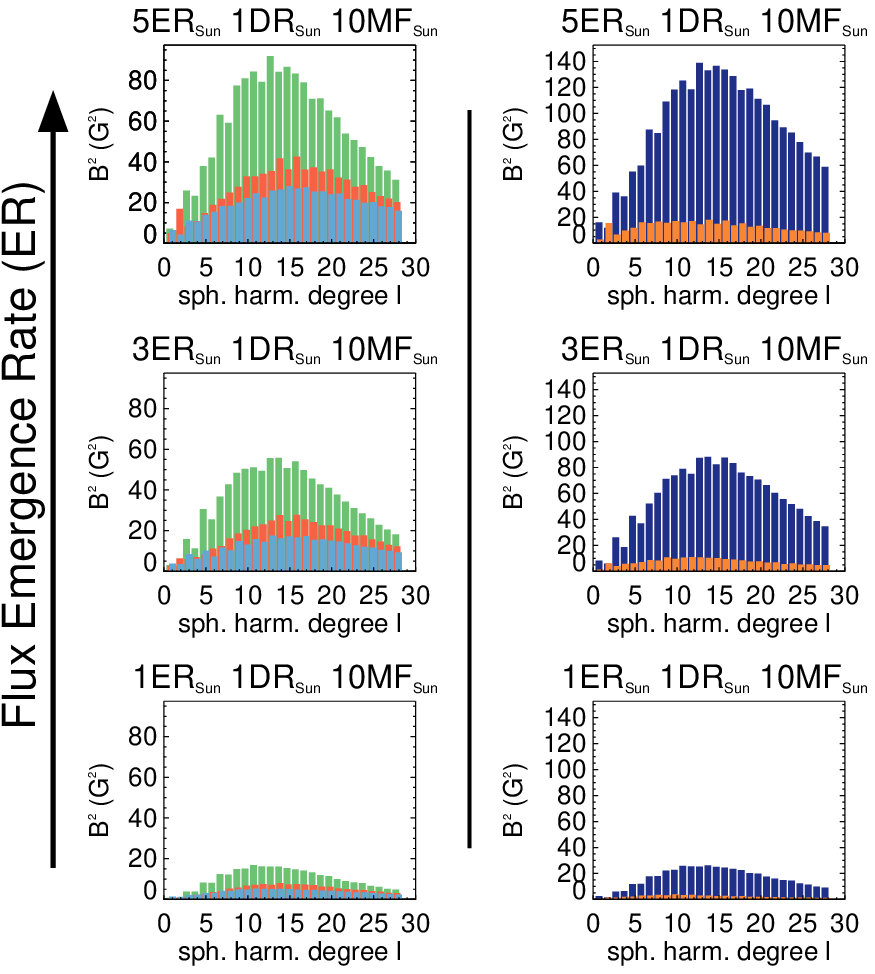}
  \caption{The magnetic energy distributions $\langle B^2\rangle(\ell)$ for the higher meridional flow simulations. Provides a direct comparison with the \textit{middle columns} of Fig.~\ref{fig:BP_E_RadAziMer} and \ref{fig:BP_E_TorPol} respectively. The same format as in Fig.~\ref{fig:BP_E_RadAziMer} and \ref{fig:BP_E_TorPol} is used.}
  \label{fig:BP_E_RadAziMerPolTor_MF}
\end{figure*}

Figure~\ref{fig:BP_E_RadAziMerPolTor_MF} displays the energy distributions $\langle B^2\rangle(\ell)$ for simulations with higher meridional flow. The \textit{left column} shows the radial $\langle B^2_{\mathrm{rad}}\rangle(\ell)$ (green bars), azimuthal $\langle B^2_{\mathrm{azi}}\rangle(\ell)$ (red bars) and meridional $\langle B^2_{\mathrm{mer}}\rangle(\ell)$ (blue bars) energy distribution for a direct comparison with the \textit{middle column} of Fig.~\ref{fig:BP_E_RadAziMer}. The \textit{right column} displays the poloidal $\langle B^2_{\mathrm{pol}}\rangle(\ell)$ (plum bars) and toroidal $\langle B^2_{\mathrm{tor}}\rangle(\ell)$  energy distribution (orange bars) for a direct comparison with the \textit{middle column} of Fig.~\ref{fig:BP_E_TorPol}.

\begin{figure}
  \includegraphics[width=0.55\columnwidth]{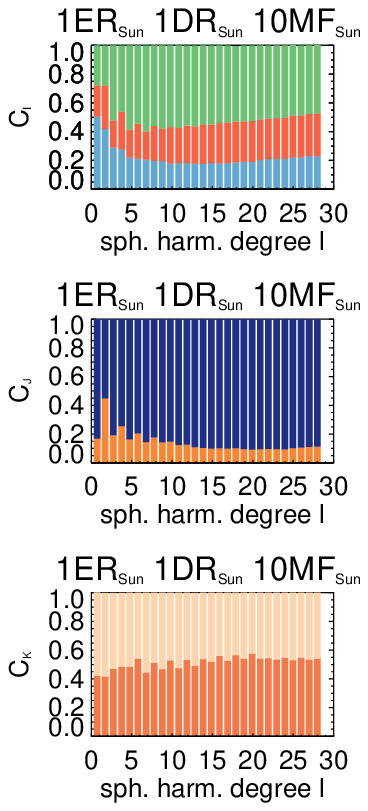}
  \caption{The cumulative total $C(\ell) = \sum f(\ell)$ of the fractions for the different field components for a higher meridional flow simulation. Provides a direct comparison with the simulated Sun, see Fig.~\ref{fig:BP_f_All} \textit{second column}. The same format as in Fig.~\ref{fig:BP_f_All} is used.} 
  \label{fig:BP_f_All_MF}
\end{figure}

Figure~\ref{fig:BP_f_All_MF} show the cumulative total $C(\ell) = \sum f(\ell)$ of the different magnetic field fractions for the higher meridional flow simulation with solar differential rotation and flux emergence rate for a direct comparison with the \textit{second column} of Fig.\ref{fig:BP_f_All}. The same format as in Fig.~\ref{fig:BP_f_All} is used.


\bsp	
\label{lastpage}
\end{document}